\documentclass[reprint,superscriptaddress,aps] {revtex4-2}

\usepackage{graphicx}
\usepackage{hyperref}
\hypersetup{hidelinks}
\usepackage{color}
\usepackage{autobreak}
\usepackage{dcolumn}
\usepackage{verbatim}

\usepackage{xcolor}
\usepackage{xpatch}
 \usepackage{amsmath}
\makeatletter

\begin{document}
\preprint{APS/123-QED}

\title{Strong phase correlation between diffusons governs heat conduction in amorphous silicon}

\author{Zhongwei Zhang}
\email{zhongwei@iis.u-tokyo.ac.jp}
\affiliation{Institute of Industrial Science, The University of Tokyo, Tokyo 153-8505, Japan} 
 
\author{Yangyu Guo}
\affiliation{Institute of Industrial Science, The University of Tokyo, Tokyo 153-8505, Japan}

\author{Marc Bescond}
\affiliation{Laboratory for Integrated Micro and Mechatronic Systems, CNRS-IIS UMI 2820, The University of Tokyo, Tokyo 153-8505, Japan}
\affiliation{IM2NP, UMR CNRS 7334, Aix-Marseille Université, Faculté des Sciences de Saint Jérôme, Case 142, 13397 Marseille Cedex 20, France}

\author{Jie Chen}
\email{jie@tongji.edu.cn}
\affiliation{Center for Phononics and Thermal Energy Science, School of Physics Science and Engineering and China-EU Joint Lab for Nanophononics, Tongji University, Shanghai 200092, People's Republic of China}

\author{Masahiro Nomura}
\email{nomura@iis.u-tokyo.ac.jp}
\affiliation{Institute of Industrial Science, The University of Tokyo, Tokyo 153-8505, Japan}

\author{Sebastian Volz}
\email{volz@iis.u-tokyo.ac.jp}
\affiliation{Institute of Industrial Science, The University of Tokyo, Tokyo 153-8505, Japan}
\affiliation{Laboratory for Integrated Micro and Mechatronic Systems, CNRS-IIS UMI 2820, The University of Tokyo, Tokyo 153-8505, Japan}
\affiliation{Center for Phononics and Thermal Energy Science, School of Physics Science and Engineering and China-EU Joint Lab for Nanophononics, Tongji University, Shanghai 200092, People's Republic of China}

\date{\today}

\begin{abstract}

Understanding the thermal vibrations and thermal transport in amorphous materials is an important but long-standing issue in several theoretical and practical fields. Using direct molecular dynamic simulations, we demonstrate that the strong phase correlation between 
local and non-propagating modes, commonly named diffusons in the terminology of amorphous systems, triggers conduction of heat. 
By considering the predominance of collective excitations in amorphous silicon, the predominant contribution of diffusons, due to phase correlation, is predicted, which further reveals the unique temperature and length dependences of thermal conductivity in amorphous silicon. The explored wavelike behavior of diffusons uncovers the unsolved physical picture of mode correlation in prevailing models and further provides an interpretation of their ability to transport heat. This work introduces a framework for understanding thermal vibrations and thermal transport in amorphous materials, as well as perspectives on the wave nature of thermal vibrations.

\end{abstract}

\pacs{Valid PACS appear here}
\maketitle

\section{Introduction}

Amorphous matter is one of the fundamental states of materials in nature and laboratories. While an adequate understanding of phonons in crystals is well established, a debate still exists regarding the physical pictures of thermal vibrations and of thermal transport in amorphous materials \cite{RevModPhys.74.991,Cahill1987,Wingert2016,DeAngelis2019,Zhou2020,Zhang2020}. Due to the loss of the long-range lattice periodicity, the concept of phonons becomes invalid in amorphous materials and the application of the phonon-gas model for thermal transport is accordingly failing \cite{Allen1993,Allen1993a,Allen1999a,Lv2016,Lv2016-2}. In the past decades, many attempts have proposed theoretical frames to describe thermal vibrations and transport in amorphous materials \cite{Allen1993,Allen1993a,Allen1999a,Simoncelli2019,Isaeva2019a,Xi2017,Zhou2018a}. Particularly, Allen and Feldman \cite{Allen1993a,Allen1993,Bickham1999} established a classification of lattice vibrations into propagons, diffusons and locons. Propagons are low-frequency propagating waves and locons are high frequency localized states, while diffusons are local modes characterized by intermediate frequencies. Their further studies demonstrated that the off-diagonal terms of the group velocity operator plays critical roles in the transport of diffusons \cite{R.J.Hardy1968,Allen1993a,Allen1999a}. In a recent work, Isaeva $et$ $al.$ \cite{Isaeva2019a} developed a quasi-harmonic Green-Kubo (QHGK) model to study the thermal transport in amorphous materials by including the correlation of scatterings between different modes. The Allen-Feldman and QHGK models share the common conceptual basis of mode correlation. However, how the correlation impacts on thermal transport and its physical representation remain unclear.

On the other hand, the study of the physical picture of thermal vibrations in amorphous materials also attracts significant research attention. In both Allen-Feldman and QHGK models, thermal vibrations are treated as plane waves in which the normal modes and scatterings are obtained from the harmonic and anharmonic lattice dynamic approaches, respectively \cite{Bickham1999,Allen1999a,Isaeva2019a}. The utilized normal modes and group velocity, which are both ill-defined in amorphous solids \cite{Lv2016}, raise questions on the fundamentals of Allen-Feldman and QHGK descriptions. In addition, in analogy to liquids, the collective excitations are experimentally observed and applied to understand the thermal vibrations in amorphous materials \cite{Shintani2008,Larkin2014,Moon2018b,Moon2019,Larklin2020,Kim2021}. Moon $et$ $al.$ \cite{Moon2018b,Moon2019} found that the thermal transport in amorphous silicon (a-Si) is dominated by acoustic collective excitations rather than by normal modes and that the elastic scattering predominates the scattering processes. The variation of lifetime obtained from collective excitations is well consistent with the frequency of the boson peak and the Ioffe-Regel crossover \cite{Shintani2008,Moon2019}. A theoretical model that can assess the thermal transport in amorphous materials by incorporating the collective excitations, however, is still missing. In addition, the controversies regarding the dependencies of thermal conductivity ($\kappa$) on temperature and length remain topics of debate \cite{Shenogin2009,Park2014,Saaskilahti2016a,Kim2021,Zhou2021}.

In this work, we theoretically investigate thermal transport in amorphous silicon, in which thermal vibrations are simultaneously interpreted as the composition of particlelike and wavelike components. The strong phase correlation or in other terms, the coherence between several diffusons, is firstly studied from wave-packet simulations. The distinct behaviors between propagons and diffusons are specifically discussed. Then, we apply our recently developed coherence heat conduction model \cite{zhang2021heat} by including lifetimes and coherence times. The significant contribution of phase correlation to thermal conductivity is deeply discussed. We finally demonstrate a fundamental, unrevealed and remarkably accurate perspective to understand thermal vibrations and transport in amorphous materials.

\section{Methodology}

\subsection{Molecular dynamic simulations}

All molecular dynamic (MD) simulations are carried out using the LAMMPS package \cite{Plimpton1995} with a time step of 0.35 fs. The Si-Si interactions in silicon systems are modeled by the Stillinger-Weber potential \cite{Stillinger1985}. a-Si is prepared from a melt-quench procedure. The details about this procedure can be found in Ref. \cite{France_Lanord_2014}. In this work, the a-Si system contains 4096 atoms and is studied in the equilibrium molecular dynamics (EMD) simulations with periodic boundary conditions in all directions. After the structure relaxation and thermal equilibration in the isothermal-isobaric (NPT) ensemble for 500 ps, EMD simulations with the microcanonical (NVE) ensemble are performed to record the atomic trajectories.

\subsection{Dynamical structure factor}

By treating the thermal vibrations as collective excitations, the vibrations in amorphous materials can be studied from the dynamical structure factor (DSF). The longitudinal and transverse DSF, $S_{L}\left ( \mathbf{q },\omega \right )$ and $S_{T}\left ( \mathbf{q },\omega \right )$, are respectively given as \cite{Shintani2008,boon1991}

\begin{eqnarray}
S_{\alpha }\left ( \mathbf{q },\omega \right )=\frac{\left |\mathbf{q }  \right |^{2}}{2\pi \omega ^{2}}\int dt\left \langle \mathbf{u}_{\alpha}\left ( \mathbf{q },t \right )\cdot \mathbf{u}_{\alpha}\left ( \mathbf{q },0 \right ) \right \rangle e^{i\omega t},
\label{eq1}
\end{eqnarray}

\noindent where $\alpha $ can be $L$ (longitudinal) or $T$ (transverse) polarization. $\mathbf{q }$ refers to the wavevector and $\omega$ to the angular frequency. The time $t$ dependent collective velocity $\mathbf{u}_{\alpha}$ can be further expressed as

\begin{eqnarray}
\mathbf{u}_{L}\left ( \mathbf{q },t \right )=\frac{1}{N}\sum_{i} \left ( \mathbf{v}_{i}\left ( t \right ) \cdot \mathbf{\hat{q}  }\right )\mathbf{\hat{q}  } e^{i\mathbf{q }\cdot \mathbf{r}_{i}\left ( t \right )},
\label{eq2}
\end{eqnarray}

\begin{eqnarray}
\mathbf{u}_{T}\left ( \mathbf{q },t \right )=\frac{1}{N}\sum_{i} \left (  \mathbf{v}_{i}\left ( t \right )- \left ( \mathbf{v}_{i}\left ( t \right ) \cdot \mathbf{\hat{q}  }\right )\mathbf{\hat{q}  }\right )e^{i\mathbf{q }\cdot \mathbf{r}_{i}\left ( t \right )},
\label{eq3}
\end{eqnarray}

\noindent where $\mathbf{\hat{q} }=\mathbf{q }/\left | \mathbf{q } \right |$. $\mathbf{v}_{i}\left ( t \right )$ and $\mathbf{r}_{i}\left ( t \right )$ are respectively the atomic velocity and position of the $i$-th atom at time $t$. $N$ denotes the total number of atoms. The atomic information of $\mathbf{v}_{i}\left ( t \right )$ and $\mathbf{r}_{i}\left ( t \right )$ can be calculated from the MD simulations of the a-Si system.

\subsection{Green-Kubo approach}

During the EMD simulations, the thermal conductivity of amorphous silicon is also calculated based on the Green-Kubo approach from the autocorrelation of the heat flux $\mathbf{J}\left ( t \right )$ \cite{Kubo1957,Kubo1957a,Volz2000-1}

\begin{eqnarray}
\kappa =\frac{V}{3k_{B}T^{2}}\int \left \langle \mathbf{J}\left ( t \right ) \cdot  \mathbf{J}\left ( 0\right )\right \rangle dt,
\label{eqs1}
\end{eqnarray}

\noindent where $V$ corresponds to the system volume, $k_{B}$ is the Boltzmann constant and $T$ the temperature. The heat flux is calculated as \cite{Torii2008}

\begin{eqnarray}
\mathbf{J}=\frac{1}{V}\left [ \sum_{i}^{N}E_{i}\mathbf{v} _{i} +\frac{1}{2}\sum_{i}^{N}\sum_{j> i}^{N}\left ( \mathbf{F}_{ij}\cdot \left ( \mathbf{v} _{i}+\mathbf{v} _{j} \right ) \right )\mathbf{r}_{ij}\right ],
\label{eqs1}
\end{eqnarray}

\noindent where $E_{i}$ is the total energy of the $i$-th atom. $\mathbf{r}_{ij}$ and $\mathbf{F}_{ij}$ refer to the distance and force between two atoms $i$ and $j$, respectively. For each case, 10 independent runs are performed in order to obtain the ensemble-averaged of $\kappa$. The correlation time considered in our simulation is long enough to ensure the proper decay of the heat current autocorrelation function.

\subsection{Wave-packet simulation}

The wave-packet simulation applied to study the propagation of propagons and diffusons is implemented by altering the atomic position from the equilibrium state. To ensure enough space and time for the observation of wave-packet propagation along one direction, a 20$\times$1$\times$1 supercell based on a system that contains 4096 atoms is applied in the wave-packet simulations. Firstly, the structure is relaxed at 0.01 K in the canonical (NVT) ensemble for 300 ps. At such a low temperature, we can ensure that the system is in a weakly anharmonic state and inelastic scattering is not involved. Then, the space dependent atomic displacement ($\Delta D $) of the wave-packet is introduced in the following form \cite{Schelling2002,Hu2018,Jiang2021}

\begin{eqnarray}
\Delta D \left ( x \right )=A_{0}e^{\left [ i \left ( x-x_{0} \right )/\lambda \right ]}e^{\left [ -\frac{1}{2}\left ( \frac{x-x_{0}}{\Delta} \right) ^{2} \right ]},
\label{eqs1}
\end{eqnarray}

\noindent where $A_{0}$ denotes the amplitude and $x_{0}$ the center position of the wave-packet. $\lambda$ is the wavelength of a specific mode and $\Delta$ is the spatial width of the wave-packet. The relationship between wavelength $\lambda$ and wavevector $\mathbf{q }$ is $\lambda =\frac{2\pi }{\left | \mathbf{q } \right |}$. In order to introduce several wave-packets, $A_{0}$ and $x_{0}$ are randomly set to generate a correspondingly random distribution of wave-packets.

\subsection{Coherence calculation}

\subsubsection{Wavelet transform approach}

The temporal coherence of thermal vibrations can be defined in the following basis \cite{Zhang2021}

\begin{eqnarray}
\centering
\psi_{\omega,t_{0},\Delta_{t} } \left ( t \right )=\pi ^{-\frac{1}{4}}\Delta_{t} ^{-\frac{1}{2}}e^{\left [ i\omega \left ( t-t_{0} \right ) \right ]}e^{\left [ -\frac{1}{2}\left ( \frac{t-t_{0}}{\Delta_{t}} \right) ^{2} \right ]},
\label{eqs2}
\end{eqnarray}

\noindent where $\Delta_{t} $ defines the temporal duration of wave-packet. $t$ corresponds to the time variable, and $t_{0}$ to the position of highest amplitude in the wave-packet and also corresponds to the time evolution in the wavelet space. Inside the wave-packet, plane waves are in phase, so the $\Delta_{t}$ term in Eq. (\ref{eqs2}) is a measure of the temporal coherence of thermal phonons. Here, we define the wave-packet full-width at half-maximum (FWHM) as the coherence time $\tau^{c}=2\sqrt{2ln2}\Delta_{t}$. The temporal coherence information of thermal vibrations can be calculated from the wavelet transform as \cite{Shiomi2006,Baker2012,Zhang2021}

\begin{eqnarray}
\Lambda \left ( \omega,t_{0},\tau^{c}\right )=\int \psi_{\omega,t_{0},\tau^{c} } \left ( t \right )F\left (  t\right )dt,
\label{eqs3}
\end{eqnarray}

\noindent where $F\left (  t\right )$ denotes the time dependent dynamical quantity, which is chosen as the collective velocity $\mathbf{u}_{\alpha}\left ( \mathbf{q },t \right )$ from Eqs. (\ref{eq2}) and (\ref{eq3}).

Besides the temporal coherence, the wavelet transform is also applied to calculate the spatial coherence, i.e., the spatial extension of wave-packets. By replacing the frequency and the temporal quantities in Eqs. (\ref{eqs2}) and (\ref{eqs3}) by wavelength and spatial quantities respectively, the basis function is obtained as follows

 \begin{eqnarray}
\centering
\psi_{\lambda ,x_{0},\Delta_{x} } \left ( x \right )=\pi ^{-\frac{1}{4}}\Delta_{\lambda} ^{-\frac{1}{2}}e^{\left [ i \left ( x-x_{0} \right )/\lambda \right ]}e^{\left [ -\frac{1}{2}\left ( \frac{x-x_{0}}{\Delta_{x}} \right) ^{2} \right ]},
\label{eqs4}
\end{eqnarray}

\noindent and

 \begin{eqnarray}
\Lambda '\left ( \lambda,x_{0},l ^{c}\right )=\int \psi_{\lambda,x_{0},l ^{c} } \left ( t \right )F'\left (  x\right )dt.
\label{eqs5}
\end{eqnarray}

\noindent $F'\left (  x\right )$ refers to the space dependent quantity, which is the atomic velocity averaged over $y$ and $z$ directions. $l^{c}$ denotes the coherence length obtained from $l^{c}=2\sqrt{2ln2}\Delta_{x}$.

\subsubsection{Vibration decay}

The wave-packets are distributed along the coherence time by using Eq. (\ref{eqs3}) and can be further investigated by building the time-averaged number density versus coherence time 

\begin{eqnarray}
D\left (\omega ,\tau ^{c} \right )=\frac{1}{N_{t_{0}}}\sum_{t_{0}}\frac{N\left ( \omega,t_{0},\tau^{c} \right )}{\sum_{\tau ^{c}}N\left ( \omega,t_{0},\tau ^{c} \right )},
\label{eqs5}
\end{eqnarray}

\noindent where $N_{t_{0}}$ denotes the number of terms in the sum. At the frequency $\omega$, the time dependent number of a given coherence time $N\left ( t_{0},\tau ^{c}  \right )$, here called number density, can be calculated as $N\left ( t_{0},\tau ^{c}  \right )=\frac{1}{2}m\left | \Lambda \left ( \omega,t_{0},\tau ^{c}  \right ) \right |^{2}/\hbar\omega$. 
 Then, the vibration decay can be calculated from the correlation function

\begin{eqnarray}
Cor\left ( t,\tau ^c \right )=\frac{\left \langle {\Delta N}\left (t,\tau ^c   \right ) {\Delta N}\left (0,\tau ^c \right ) \right \rangle}{\left \langle {\Delta N}\left (0,\tau ^c \right ) {\Delta N}\left (0,\tau ^c \right ) \right \rangle},
\label{eqs6}
\end{eqnarray}

\noindent here ${\Delta N}(t,\tau ^c)=N\left ( t,\tau ^c \right )-\left \langle N(t,\tau ^c) \right \rangle_{t}$. To obtain a correlation function $Cor\left ( t \right )$ independent of the coherence time, an average over coherence times can be further implemented

\begin{eqnarray}
Cor\left ( t \right )=\sum_{\tau ^{c}}D\left (\tau ^{c} \right )Cor\left ( t,\tau ^c \right ).
\label{eqs7}
\end{eqnarray}

\subsubsection{Coherence time and Lifetime}

The vibration decays are as numerous as the number of coherence times. As input to the general thermal conductivity expression, we define the averaged coherence time ($\bar{\tau}^{c}$) 

\begin{eqnarray}
\bar{\tau}^{c}=\sum_{\tau ^{c}}D\left (\tau^{c} \right )\tau^{c}.
\label{eqs8}
\end{eqnarray}

Mean lifetimes ($\bar{\tau}^{p}$) are obtained by fitting the averaged phonon decay $Cor\left ( t \right )$ as \cite{Zhang2021}
  
\begin{eqnarray}
Cor\left ( t \right )= e^{-\frac{t}{2\bar{\tau}^{p} }}e^{-4ln2\frac{t^{2}}{\bar{\tau}^{c2}}}.
\label{eqs9}
\end{eqnarray}

In the following sections, $\tau ^{p}$ and $\tau^{c}$ refer to $\bar{\tau} ^{p}$ and $\bar{\tau} ^{c}$ of the above formulas, respectively.


\section{DSF spectrum}

\begin{figure}[t]
\includegraphics[width=1.0\linewidth]{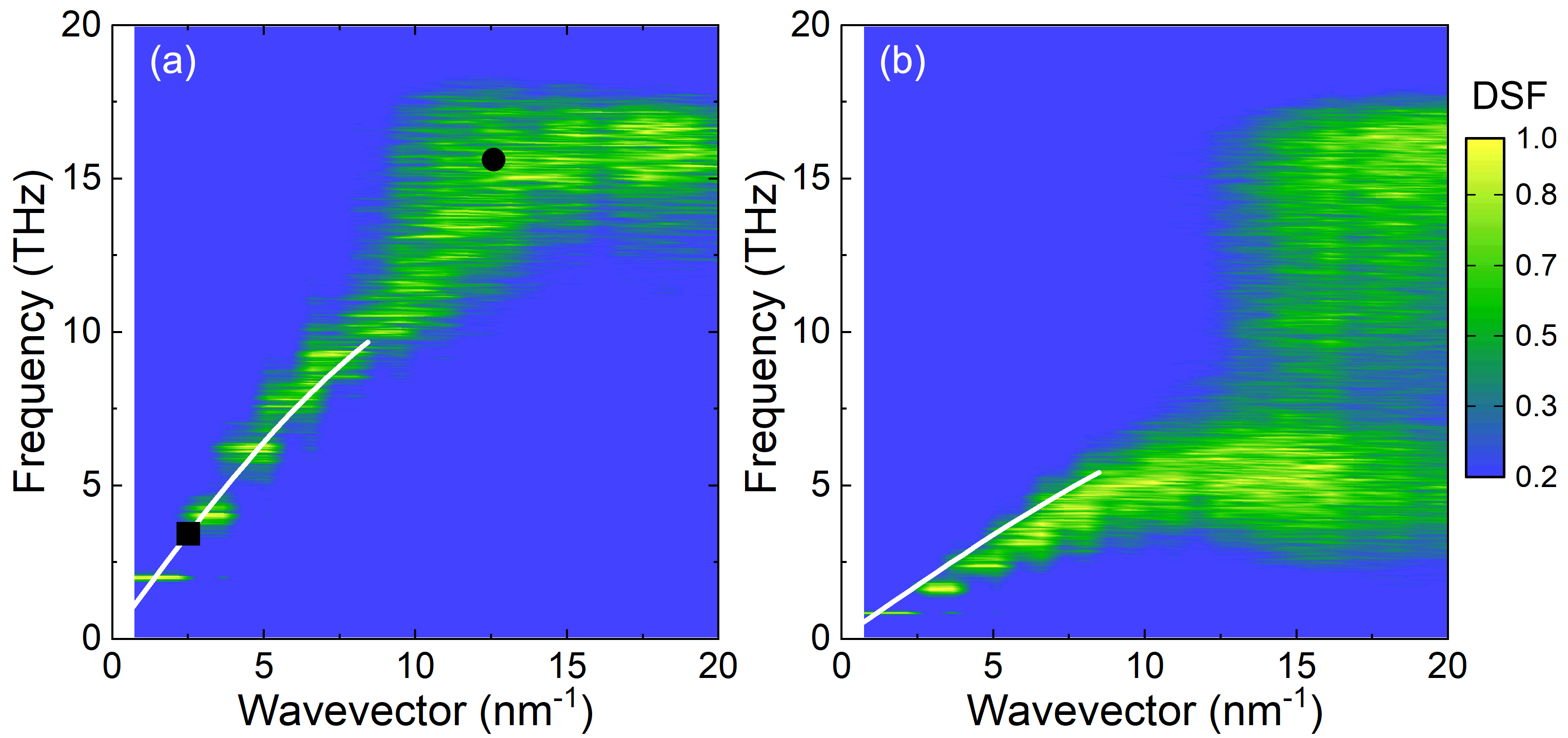}
\caption{The calculated dynamical structure factor (DSF) of amorphous silicon for the longitudinal (a) and transverse (b) excitations at room-temperature. The calculated amorphous silicon contains 4096 atoms. The solid white lines are respectively the phonon dispersion of the longitudinal branch (a) and the transverse branch (b) for crystal silicon. The dots in (a) are the specific vibrations that applied in wave-packet simulations.
}
\label{figure1}
\end{figure}

Considering the atomic instability and the absence of lattice periodicity in amorphous materials, thermal vibrations in a-Si are treated as collective excitations as studied from the DSF \cite{Moon2019,Kim2021}, which also refer to acoustic collective excitations, where atoms are vibrating along the same polarization (See Eqs. (\ref{eq2}) and (\ref{eq3})). The concept of collective excitation was initially implemented to describe liquids or liquid-like matters \cite{boon1991}. The similarity in the lattice aperiodicity and atomic instability makes the collective excitation a reasonable description of thermal vibrations in amorphous materials as discussed before in Refs. \cite{Moon2019,Kim2021,moon2021examining}. The calculation method of DSF is presented in the Methodology section. The calculated DSF is shown in Fig. \,\ref{figure1}, which includes the longitudinal and transverse polarizations. 

Similarly to the behavior of phonons in crystals, the longitudinal and transverse polarizations show distinct tendencies due to a lower group velocity of the collective transverse mode. In addition, the results in Fig. \,\ref{figure1} show that the dispersion of collective excitations in the small wavevector region agrees well with the acoustic branches of silicon crystal (white lines), indicating that the collective excitations can highly inherit thermal vibrations from the corresponding crystal. Moreover, there is an obvious increase in the broadening of polarizations after $\sim $ 10 nm$^{-1}$. The threshold wavevector of this broadening corresponds to the first Brillouin zone boundary of crystal and the transition from propagons to diffusons. Figure \,\ref{figure1} indicates that the longitudinal and transverse polarizations have different transition frequencies. Note that the transition frequency of the transverse polarization is directly linked to the frequency of the boson peak \cite{Shintani2008}. As demonstrated before \cite{Moon2019}, the broadening is inversely related to the modal lifetime. The transition from propagons to diffusons is appearing as a significant change of lifetime, which will be highlighted in Sec. VI.

 \begin{figure*}[t]
\includegraphics[width=0.8\linewidth]{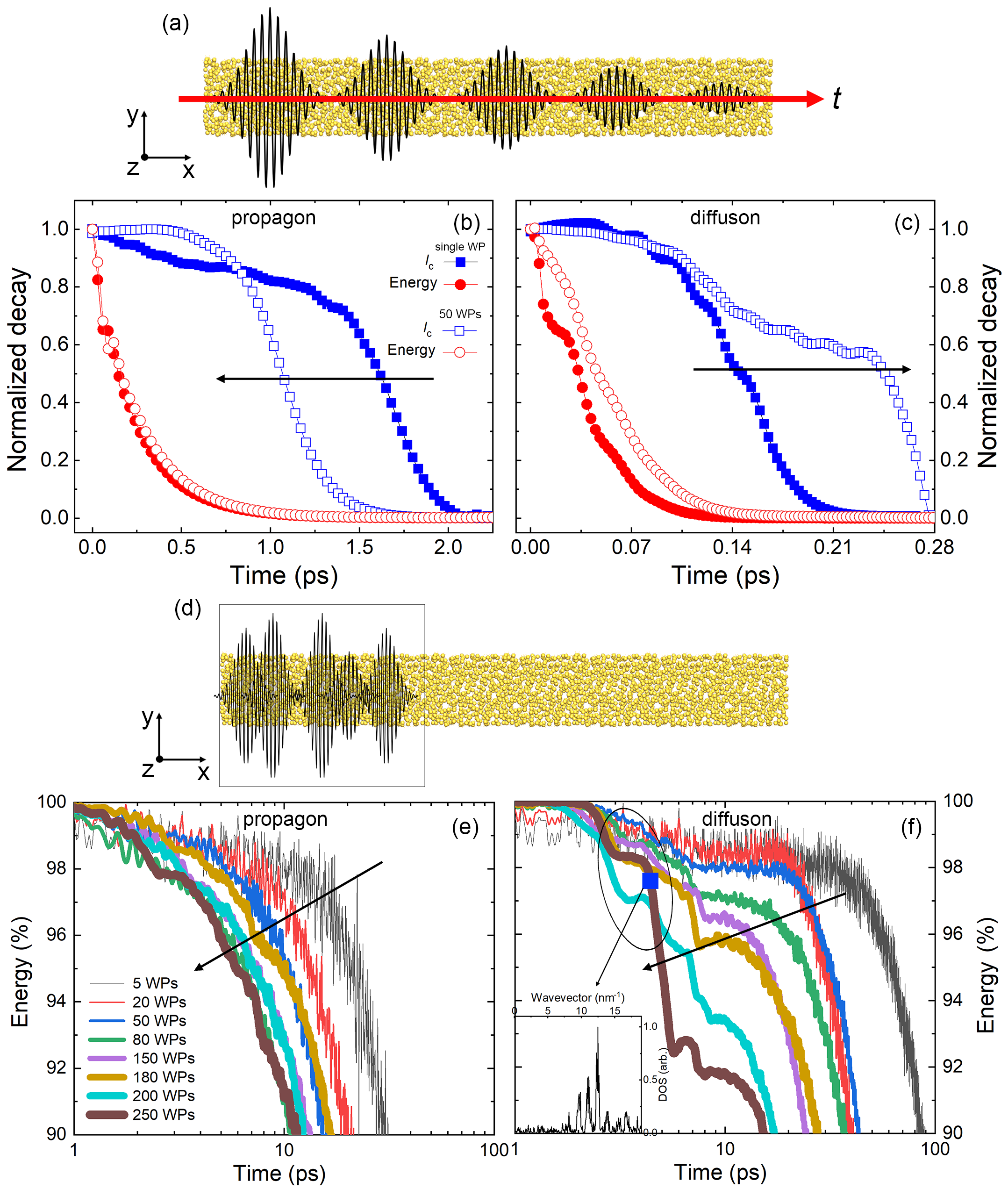}
\caption{(a) Schematic figure of wave-packets propagating in amorphous silicon. (b) Normalized decay trends of energy and coherence length ($l_{c}$) versus simulation time in the cases of single propagon wave-packet and of 50 propagon wave-packets. (c) Normalized decay trends of energy and coherence length ($l_{c}$) versus simulation time in the cases of a single diffuson wave-packet and of 50 diffuson wave-packets. The arrows indicate the shift of $l_{c}$ with increasing wave-packets number. (d) Schematic of wave-packet excitations in amorphous silicon. The black box indicates the initial excited region with a length of 30 nm and the corresponding energy is recorded. Energy percentage versus simulation time for the different numbers of (e) propagon wave-packets and (f) diffuson wave-packets. The arrows indicate the shift of energy decrease with increasing wave-packets number. The inset in (f) shows the density of states (DOS) at 4 ps in the 250 wave-packets case.
}
\label{figure2}
\end{figure*}

\section{Wave-packet simulation}

In contrast to the estimation of the frequencies of the lattice vibrations obtained from the normal mode analysis, which is only carried out at the Gamma point, the DSF in Fig. \,\ref{figure1} can also provide the information of wavevector for different vibrations but in the frame of collective excitations as proposed in Refs. \cite{Moon2019,Kim2021,moon2021examining}. By implementing the wave-packet simulation, we further study the detailed transport properties of propagons and diffusons. The approach to excite wave-packets is presented in Sec. II. D, where the environment temperature is set to 0.01 K to avoid inelastic/anharmonic scattering. To excite different states in a-Si, the wavevectors for longitudinal propagons and diffusons are respectively chosen as 2.5 nm$^{-1}$ and 12.5 nm$^{-1}$ as marked by the dots in Fig. \,\ref{figure1}(a). Figure \,\ref{figure2}(a) depicts the schematic figure of a wave-packet simulation and its propagation, in which the excited wave-packet is propagating along the $x$ direction in a-Si. In each interval of 0.05 ps, the energy of this propagating wave-packet is recorded and the spatial wavelet transform is performed to calculate the variation of the coherence length ($l_{c}$) of this wave-packet (See Sec. II. E). Figures \,\ref{figure2}(b) and \,\ref{figure2}(c) respectively show the variation of energy and $l_{c}$ of propagons and diffusons. The energies of these two types of vibrations gradually decay during their propagation. As the environment temperature is low enough to weaken the anharmonic scattering, the energy decrease in Figs. \,\ref{figure2}(b) and \,\ref{figure2}(c) indicates the significance of elastic scattering in the dissipation of thermal vibrations in amorphous materials. In addition, the energy of diffusons decays much faster than that of propagons indicating a stronger elastic scattering for the less-propagating diffusons, which agrees well with previous predictions \cite{Moon2019,Kim2021}.

On the other hand, a unique effect of elastic scattering on coherence is observed. Figures \,\ref{figure2}(b) and \,\ref{figure2}(c) show that the coherence length $l_{c}$ decreases significantly slower than the energy. The difference indicates that during elastic scattering, the vibrational amplitude of wave-packetes decreases, resulting in the dissipation of energy. Simultaneously, the spatial extension of wave-packets remains constant with high coherence until the wave-packet fully disappears (See the schematic in Fig. \,\ref{figure2}(a)). To model the realistic system that contains a high density of wave-packets in the same mode, the system with 50 wave-packets is also studied. Compared to the limited effect on energy decreases due to the predominant elastic scattering, a stronger effect on $l_{c}$ is found. For propagons, $l_{c}$ decreases with the increase of the wave-packets number, which might be originated from the anharmonic interaction between wave-packets, while the $l_{c}$ of diffusons is reversely increased. The increased $l_{c}$ should resulted from the wavelike interaction between diffusons or phase correlation.

The influence of coherence on thermal transport is further investigated from the wave-packet simulations in Figs. \,\ref{figure2}(d-f). In this simulation, several wave-packets are randomly excited in a well-defined region of length 20 nm and then the energy within this region is recorded (See the box in Fig. \,\ref{figure2}(d)). As increasing the number of wave-packets, the recorded energy change correspondingly increases. At low density, diffusons show low thermal transport efficiency in Fig. \,\ref{figure2}(f). However, as the wave-packet number reaches 50, a sudden energy change appears around 4 ps, indicating a high transport efficiency, which becomes more pronounced as up to 250 wave-packets. The additional density of states calculation in the inset of Fig. \,\ref{figure2}(f) indicates that the system of 250 wave-packets at 4 ps is still dominated by diffusons with wavevector higher than 10 nm$^{-1}$ and the enhanced transport efficiency is still related to the behavior of diffusons. On the contrary, the transport efficiency of propagons in Fig. \,\ref{figure2}(e) follows a monotonic decreasing trend with simulation time.

 \begin{figure}[t]
\includegraphics[width=1.0\linewidth]{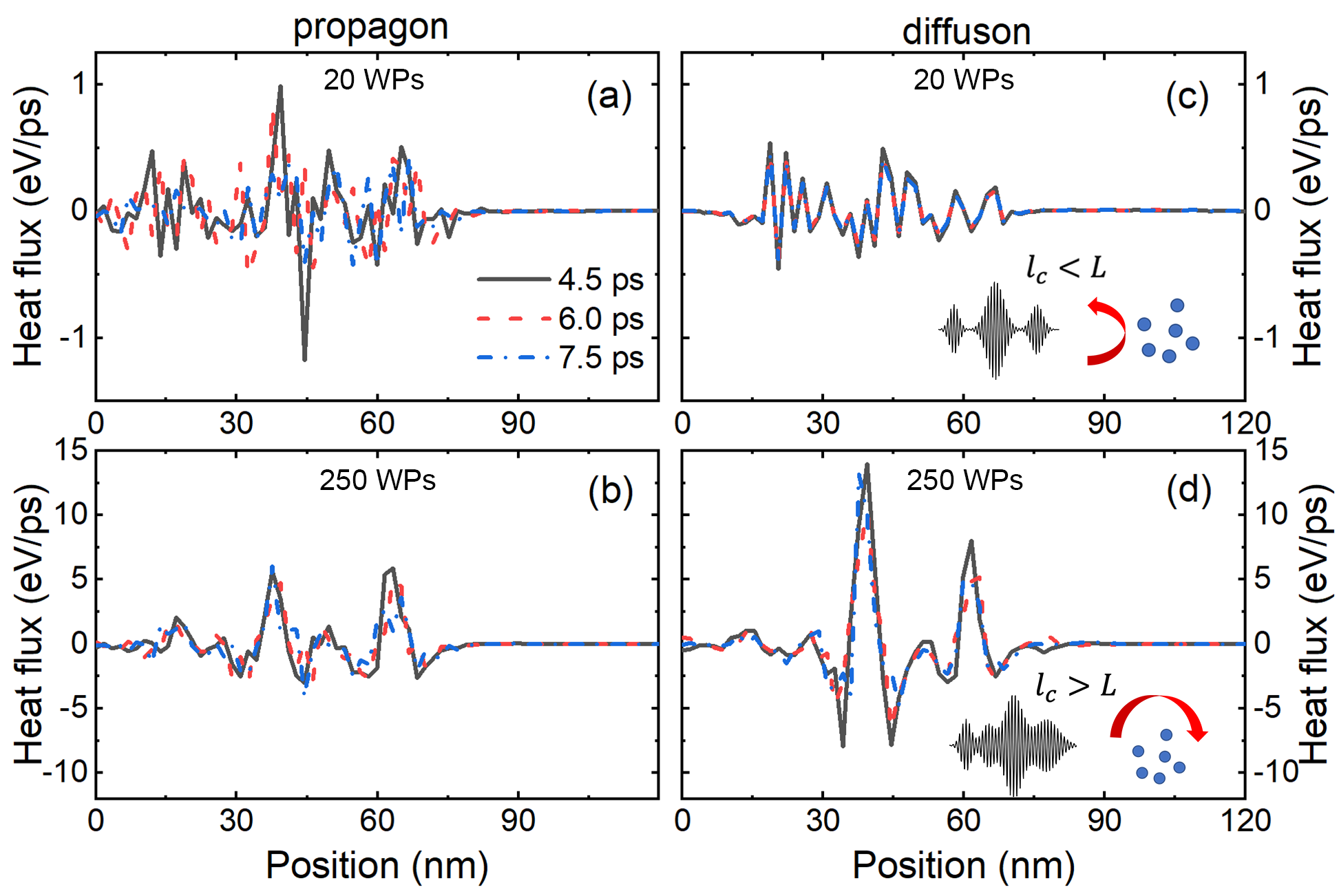}
\caption{The simulated heat flux for the cases where propagons are excited with (a) 20 wave-packets and (b) 250 wave-packets at 4 ps. The simulated heat flux for the cases where diffusons are excited with (c) 20 wave-packets and (d) 250 wave-packets at 4 ps. The inset of figure (c) shows the schematic of a low-density ensemble of wave-packets reflected by the local disorder. The inset of (d) shows the schematic of a high density ensemble of coherent diffuson wave-packets propagating through the local disorder.}
\label{figure3}
\end{figure}

\begin{figure}[b]
\includegraphics[width=1.0\linewidth]{figure4.png}
\caption{Calculated lifetime ($\tau^{p}$) and coherence time ($\tau^{c}$) of amorphous silicon at room-temperature. The lines indicate the Ioffe-Regel crossover for the transverse and longitudinal polarizations.}
\label{figure4}
\end{figure}

 \begin{figure*}[t]
\includegraphics[width=0.85\linewidth]{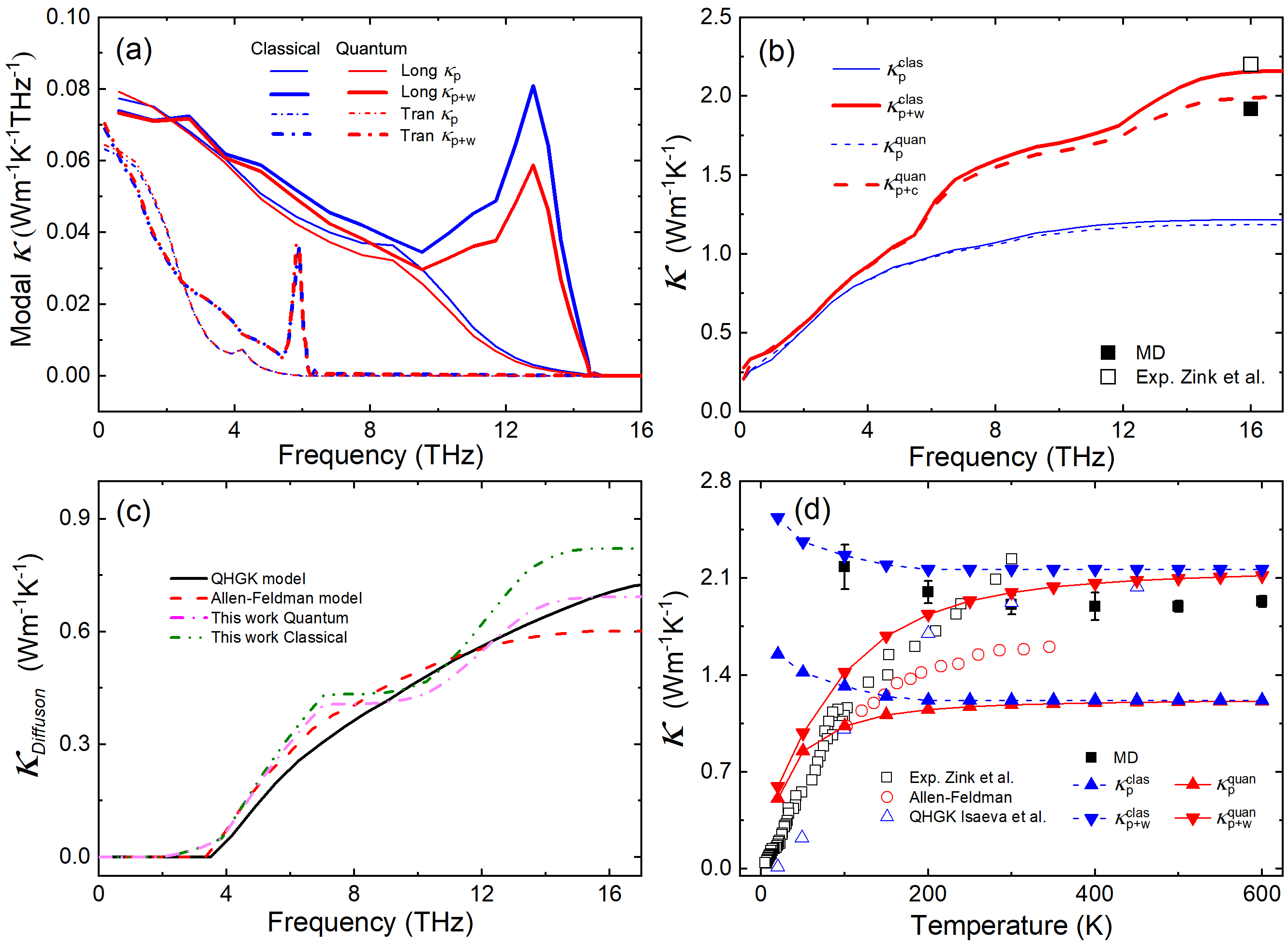}
\caption{The thermal conductivity of amorphous silicon. (a) Room-temperature modal classical and quantum thermal conductivities ($\kappa$) of amorphous silicon for the longitudinal (Long) and transverse (Tran) polarizations, respectively from pure particlelike ($\kappa_{p}$) and overall (i.e., particlelike+wavelike ($\kappa_{p+w}$)) contributions. (b) Room-temperature accumulative thermal conductivities versus frequency.  (c) Room-temperature accumulative thermal conductivities of diffusons versus frequency from the quasi-harmonic Green-Kubo (QHGK) model, the Allen-Feldman model and this work. (d) Thermal conductivities versus temperature. The symbols refer to experimental results (Exp.) \cite{Zink2006}, QHGK model, Allen-Feldman model and our molecular dynamics (MD) simulations.
}
\label{figure5}
\end{figure*}

To further evaluate the coherence effect on thermal transport efficiency, the spatial heat flux distribution is calculated in Fig. \,\ref{figure3}. The heat flux is obtained at the time of the enhanced energy change appearing $\sim$ 4 ps. For the systems with a small number of wave-packets, such as 20 wave-packets in Figs. \,\ref{figure3}(a) and 3(c), the less-propagating and low density diffusons generate a lower heat flux than in the case of propagons. While increasing the number of wave-packets to 250 in Figs. \,\ref{figure3}(b) and 3(d) where the high propagation efficiency of diffusons appears, diffusons reversely show more significant heat flux than propagons. On the other hand, compared to the dense oscillations at low density, the heat flux of diffusons at high density exhibits more unified oscillations, indicating the generation of in-phase/coherent propagating waves of diffusons due to the strong phase correlation. The phase correlation between diffusons might be related to the reported spatial correlations of local stress \cite{Gelin2016}. As displayed in the inset of Figs. \,\ref{figure3}, the less-propagating nature and low density of diffusons lead to a state of quasi-localization with low thermal transport efficiency, in which the coherence length is smaller than the length-scale of local disorder ($L$), $l_{c}<L$. However, the high density results in the less-propagating diffusons are more likely to be coherent to the wave-packets with larger extension, i.e., strongly phase correlated, making $l_{c}>L$. Consequently, a high thermal transport efficiency emerges. In a realistic system, vibrations are randomly and extensively excited by thermal energy, which is much analogous to the simulated cases with large density wave-packets (See Figs. \,\ref{figure2}(c) and \,\ref{figure3}(d)). This high transport efficiency of diffusons here agrees well with the previously reported high contribution from diffusons to thermal conductivity in amorphous materials \cite{Zhou2016b,Lv2016,Zhou2017b}. The study in Fig. \,\ref{figure3} reports the unexpected effect and physical picture of coherence on the transport of diffusons in amorphous materials.

\section{Lifetime and Coherence time}

With the coherence calculation in Sec. II. E, we estimated the lifetimes and coherence times for different collective excitation modes in Fig. \,\ref{figure4}. Similar to the frequency dependence in crystal, the lifetimes and coherence times of a-Si decreases with frequency in the low frequency region. After several Terahertz, the lifetimes of both transverse and longitudinal polarizations approaches to zero. This trend corresponds to the Ioffe-Regel crossover as indicated by the lines in Fig. \,\ref{figure4}. Obviously, the Ioffe-Regel crossover is polarization dependent and the transverse polarization is characterized by a lower frequency of Ioffe-Regel crossover. After the Ioffe-Regel crossover, thermal vibrations in a-Si fall into the diffuson region. The previous classification of propagons and diffusons from the normal mode analysis is probably insufficient, as the polarizations are not properly distinguished \cite{Allen1993a,Beltukov2013,Zhu2016,Seyf2016}.

On the other hand, the coherence time shows different dependences on frequency. For low frequency vibrations, i.e., propagons, the coherence times also decreases with the frequency and the values of coherence time are close to that of lifetimes. At high frequencies, diffusons show longer coherence time than the lifetimes. The pronounced coherence time agrees well with our previous discussion about the strong phase correlation between less-propagating modes, i.e., diffusons.

\section{Thermal conductivity}

\subsection{Coherent thermal transport}

As demonstrated in previous sections, the wavelike behavior has significant impact on the transport and dynamics of diffusons. In our recent work, by introducing the coherence related to the temporal extension of wave-packets, the coherence effect on the modal decay trend due to their collision is studied in Eq. (\ref{eqs9}) \cite{Zhang2021,zhang2021heat}. Correspondingly, the complete thermal conductivity ($\kappa _{p+w}$) including both the particlelike ($p$) and wavelike ($w$) behaviors can be expressed as \cite{Zhang2021}

\begin{eqnarray}
\kappa _{p+w} =  \frac{1}{3}\sum_{\alpha }\sum_{\nu } C _{\mathrm{v},\nu}^{clas} \upsilon _{\nu,\alpha}^{2}\sqrt{\frac{\pi }{4ln2}}\tau _{\nu}^{c} e^{\frac{{\tau _{\nu}^{c2}} }{128ln2{\tau_{\nu}^{p2}} }}.
\label{eqs15}
\end{eqnarray}

\noindent Here, $\upsilon_{\nu,\alpha}$ denotes the group velocity of mode $\nu$ along the cartesian coordinate $\alpha$. $C _{\mathrm{v},\nu}^{clas}= k_{B}/V$ is the classical specific heat per mode. The modal coherence time and lifetime are obtained from previous coherence calculations in Fig. \,\ref{figure4}. The particlelike thermal conductivity can be calculated with the Peierls-Boltzmann formula \cite{ziman2001} as $\kappa_{p} =  \frac{1}{3}\sum_{\alpha }\sum_{\lambda} C _{\mathrm{v},\nu}^{clas} \upsilon _{\nu,\alpha}^{2} \tau_{\nu}^{p}$. Obviously, the calculation of thermal conductivity from Eq. (\ref{eqs15}) depends on the quantities of frequency, group velocity, lifetime and coherence time. In contrary to the ill-defined normal mode analysis, these quantities can be obtained from the collective excitation picture provided by the DSF spectrum with well-defined dispersion as proposed by Moon $et$ $al.$ \cite{Moon2018b,Moon2019}. Especially, the group velocity can be calculated by $\boldsymbol{\upsilon}_{\nu}=d\omega_{\nu} /d\boldsymbol{q}$. The relationship between $\omega_{\nu}$ and $\boldsymbol{q}$ is determined by the DSF spectrum for the transverse and longitudinal polarizations. Our wave-packet simulations also indicate that the predominant vibrations in amorphous materials manifest strong wavelike behavior, which can be captured by Eq. (\ref{eqs15}) from coherence time. The calculations can be further quantum corrected by replacing $C _{\mathrm{v},\nu}^{clas}$ with the quantum specific heat $C _{\mathrm{v},\nu}^{qua}$, here $C _{\mathrm{v},\nu}^{qua}=\frac{k_{B}exp\left ( x  \right )}{V}\left [\frac{x}{exp\left ( x  \right )-1} \right ]^{2} $ with $x=  \frac{\hbar \omega _{\nu} }{k_{B}T} $.
 
Figure \,\ref{figure5} shows the calculated thermal conductivities from different approaches. The particlelike contribution to thermal conductivity, i.e. $\kappa_{p}$, exhibits the same tendency as the predictions in crystals, where $\kappa_{p}$ decreases with the frequency in the low-frequency region. However, the particlelike contributed thermal conductivity is lower than the one from experimental measurements \cite{Zink2006} and direct MD Green-Kubo prediction (See the accumulative $\kappa$ in Fig. \,\ref{figure5} (b) and the Green-Kubo calculation in Sec. II.C). The modal contribution indicates that only considering the particlelike behavior of thermal vibrations, the contribution from diffusons at high frequencies cannot be captured (thinner lines). This outcome also manifests that the dynamics or transport of diffusons is mainly dominated by the wavelike behaviors, which is consistent with the results of wave-packet simulations. 

We further investigate the coherence corrected thermal conductivity from Eq. (\ref{eqs15}) in Fig. \,\ref{figure5}. Compared to the weak wavelike contribution from low-frequency propagons, the wavelike thermal conductivity predominates the heat conduction of high-frequency diffusons (thicker lines in Fig. \,\ref{figure5}(a)), indicating the strong wave correlation effect for diffusons and its significant contribution to thermal transport. The large wavelike contribution from diffusons to thermal conductivity well coincides with the enhanced propagating efficiency in wave-packet simulations described in previous sections. The accumulative thermal conductivity of a-Si in Fig. \,\ref{figure5}(b) shows that the coherence corrected room-temperature thermal conductivity agrees well with the experimentally measured value and the MD Green-Kubo $\kappa$. The correction degree, i.e., $\frac{\kappa_{p+w}-\kappa_{p}}{\kappa_{p}}\times 100\%$, reaches 66 $\%$ under quantum approximation, and the contribution is mainly from frequencies after 4 THz by diffusons. In addition, due to the high frequency of diffusons, the quantum effect of the phonon population only quantitatively influences the thermal conductivity values.

Compared to the prevailing models, we provide a physical understanding of diffuson propagation arising from the strong phase correlation. The comparison between different theories yielding the estimation of the thermal conductivity from diffusons, i.e. $\kappa_{Diffuson}$, is further presented in Fig. \,\ref{figure5}(c). The contribution from diffusons is accumulated from modes with frequencies after the Ioffe-Regel crossover. Three different models correct the thermal conductivity due to the following effects: 1) in our model, the phase correlation/coherence between diffusons; 2) in the QHGK model, the correlation between different modal scatterings/lifetimes (i.e., off-diagonal terms); 3) in the Allen-Feldman model, the off-diagonal terms of group velocity operator. Figure \,\ref{figure5}(c) shows that $\kappa_{Diffuson}$ in a wide frequency region is in excellent agreement with the results from both the Allen-Feldman and QHGK models with the same quantum population. The higher values obtained at the classical limit are due to the overestimation of the phonon population at high frequencies. The results of Allen-Feldman and QHGK models are respectively calculated from the packages of GULP \cite{GULP} and $\kappa$ALD$o$ \cite{kALDo} with the same a-Si configuration as implemented in the DSF calculations. 

The comparison between the temperature dependent thermal conductivities is displayed in Fig. \,\ref{figure5}(d). Our predicted quantum thermal conductivities, i.e. $\kappa_{p+w}^{quan}$, agree satisfactorily with the predictions from other theoretical models and experimental values in the full temperature regime, showing higher accuracy compared to the Allen-Feldman model. Compared to the high thermal conductivities acquired from direct MD simulations with over-populated modes in the classical limit, the quantum effects can be effectively treated, resulting in a thermal conductivity decrease when temperature decreases. In addition, the fast convergence of thermal conductivity and the negligible temperature dependence at high temperatures indicate the weakly anharmonic scattering process and the predominant elastic scattering in a-Si. Figure \,\ref{figure5}(d) also shows that the coherence correction to the thermal conductivity of a-Si becomes more pronounced as temperature increases, manifesting the enhanced phase correlation between modes with temperature as demonstrated recently in complex crystals \cite{ Simoncelli2019}.

\subsection{Length dependent thermal conductivity}

\begin{figure}[t]
\includegraphics[width=1.0\linewidth]{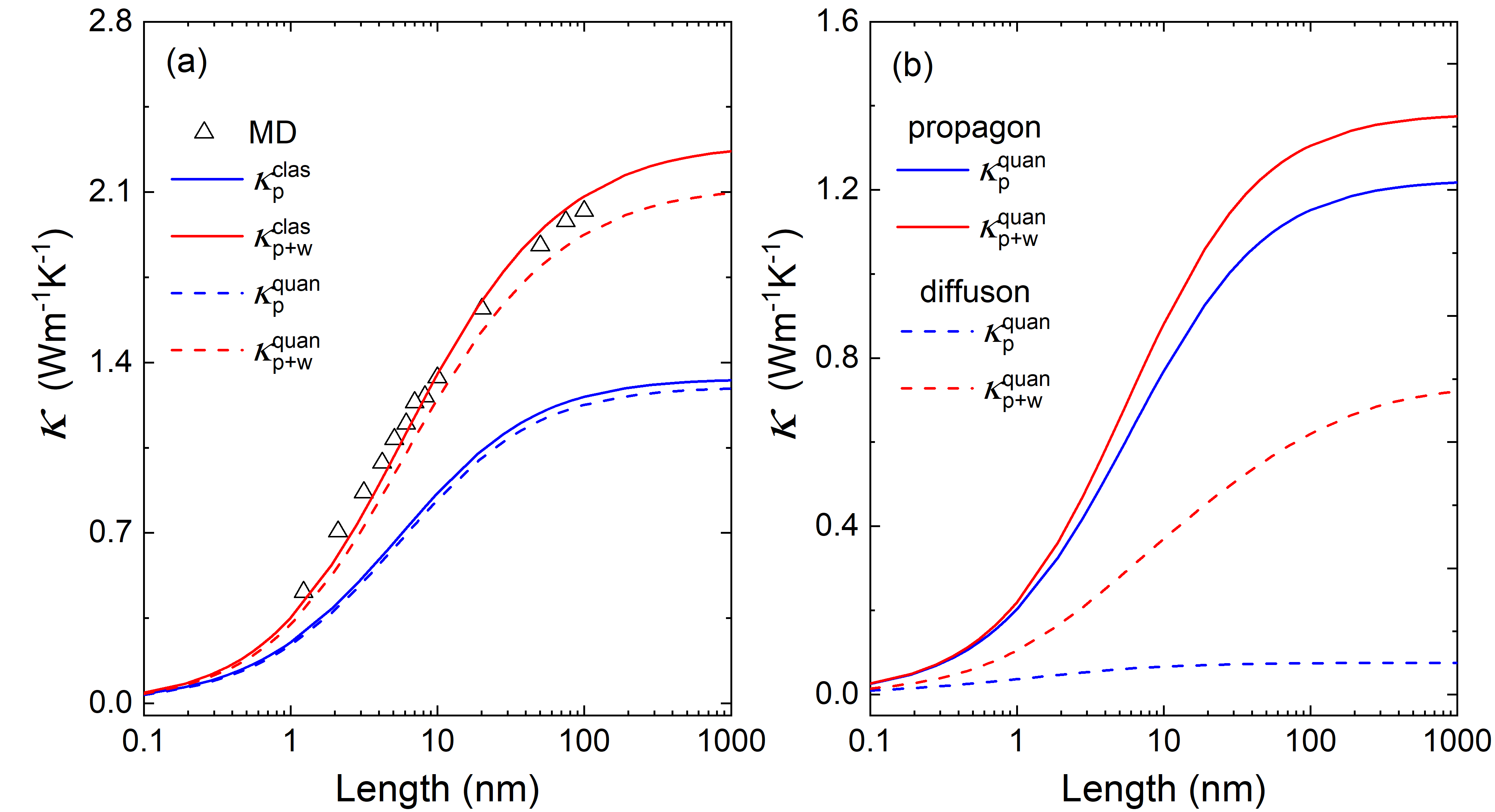}
\caption{Length-dependent room-temperature thermal conductivity of amorphous silicon. (a) Length-dependent classical and quantum thermal conductivity of amorphous silicon respectively from particlelike and wavelike contributions at room-temperature. (b) Length dependent quantum thermal conductivity respectively from propagons and diffusons at room-temperature.}
\label{figure6}
\end{figure}

The obtained vibrational properties from DSF calculations are analogous to the phononic properties of a crystal. In principle, the diverse scatterings can be incorporated into the Matthiesse's law

\begin{eqnarray}
\frac{1}{\tau _{\nu }^{tot}}=\frac{1}{\tau _{\nu }^{d}}+\frac{1}{\tau _{\nu }^{b}},
\label{eqs16}
\end{eqnarray}

\noindent where, $\tau _{\nu }^{d}$ refers to the decay time $\sqrt{\frac{\pi }{4ln2}}\tau _{\nu}^{p} e^{\frac{{\tau _{\nu}^{c2}} }{128ln2{\tau_{\nu}^{p2}} }}$ in Eq. (\ref{eqs15}) due to the anharmonic and elastic scatterings in Eq. (\ref{eqs9}). The boundary scattering time $\tau _{\nu }^{b}$ is calculated as $\tau _{\nu }^{b}=\frac{2L_{s}}{\upsilon  _{\nu }}$ \cite{Casimir1938}. $L_{s}$ is the length of the system. As a comparison, the length dependent thermal conductivity is also calculated from the non-equillrium MD simulations in Fig. \,\ref{figure6}(a) by following our previous procedure used in crystalline nanonstructures \cite{Zhang2017k}.

Figure \,\ref{figure6}(a) shows that the coherence corrected thermal conductivities are in fair agreement with MD predictions. The limited discrepancies might be originating from the uncertainty of MD simulations. The purely particlelike contribution underestimates the thermal energy carriers with long propagation lengths. Recent measurements showed that the thermal conductivity of a-Si displays a marked length dependence and the ultra-long mean free path of vibrations can reach several hundreds of nanometers \cite{Regner2013,Braun2016,Pan2020,Kim2021}. Our predictions in Fig. \,\ref{figure6}(a) indicate that this long length dependence might result from the wavelike contribution of diffusons. On the other hand, the agreement between the predictions from our model and MD simulations demonstrates the validation of treating the vibrations in amorphous materials as collective excitations that have specific polarizations and group velocities.

We further quantify the length dependent contributions into propagons and diffusons by following the Ioffe-Regel crossover. As shown in Fig. \,\ref{figure6}(b), the coherence has a limited effect on the propagon thermal conductivity. As discussed in our previous work, the small size and highly propagating wave-packet of thermal vibrations should exhibit weak intrinsic coherence \cite{Zhang2021}, which is the case of propagons in a-Si. However, diffusons, due to their less-propagating nature, can be easily correlated as waves, which yields a strong coherence. Correspondingly, the wavelike behavior results in long propagating lengths (i.e., mean free paths here) and a high diffuson thermal conductivity.

\section{Conclusion}

In this work, we demonstrate the strong phase correlation of diffusons in amorphous silicon. The wave-packet simulations show that due to their less-propagating nature and wavelike behavior, diffusons can be highly correlated and exhibit high transport efficiency. The collective excitation nature and the predominance of elastic scattering of the thermal vibrations are also demonstrated in amorphous materials. The agreements between a heat conduction model including coherence effects and other results from prevailing theories not only validate the generalization of our model in different systems but also show the unexpected and significant coherence effect on the thermal conductivity of amorphous materials. The past discrepancies and recent measurements of the temperature and length dependences of thermal conductivity are well explained by the exploration of the strong phase correlation of diffusons in amorphous silicon. This work provides a framework to understand the thermal vibrations and thermal transport in amorphous materials and reveals new perspectives in relation with the wave nature of thermal vibrations.

\section*{\label{sec:level1}Acknowledgments}

This work is partially supported by CREST JST (No. JPMJCR19I1 and JPMJCR19Q3). This research used the computational resources of the Oakforest-PACS supercomputer system, The University of Tokyo. This project is also supported in part by the grants from the National Natural Science Foundation of China (Grant Nos. 12075168 and 11890703), and Science and Technology Commission of Shanghai Municipality (Grant No. 19ZR1478600).

\bibliographystyle{apsrev4-2}
\bibliography{library}

\begin{thebibliography}{58}%
\makeatletter
\providecommand \@ifxundefined [1]{%
 \@ifx{#1\undefined}
}%
\providecommand \@ifnum [1]{%
 \ifnum #1\expandafter \@firstoftwo
 \else \expandafter \@secondoftwo
 \fi
}%
\providecommand \@ifx [1]{%
 \ifx #1\expandafter \@firstoftwo
 \else \expandafter \@secondoftwo
 \fi
}%
\providecommand \natexlab [1]{#1}%
\providecommand \enquote  [1]{``#1''}%
\providecommand \bibnamefont  [1]{#1}%
\providecommand \bibfnamefont [1]{#1}%
\providecommand \citenamefont [1]{#1}%
\providecommand \href@noop [0]{\@secondoftwo}%
\providecommand \href [0]{\begingroup \@sanitize@url \@href}%
\providecommand \@href[1]{\@@startlink{#1}\@@href}%
\providecommand \@@href[1]{\endgroup#1\@@endlink}%
\providecommand \@sanitize@url [0]{\catcode `\\12\catcode `\$12\catcode
  `\&12\catcode `\#12\catcode `\^12\catcode `\_12\catcode `\%12\relax}%
\providecommand \@@startlink[1]{}%
\providecommand \@@endlink[0]{}%
\providecommand \url  [0]{\begingroup\@sanitize@url \@url }%
\providecommand \@url [1]{\endgroup\@href {#1}{\urlprefix }}%
\providecommand \urlprefix  [0]{URL }%
\providecommand \Eprint [0]{\href }%
\providecommand \doibase [0]{https://doi.org/}%
\providecommand \selectlanguage [0]{\@gobble}%
\providecommand \bibinfo  [0]{\@secondoftwo}%
\providecommand \bibfield  [0]{\@secondoftwo}%
\providecommand \translation [1]{[#1]}%
\providecommand \BibitemOpen [0]{}%
\providecommand \bibitemStop [0]{}%
\providecommand \bibitemNoStop [0]{.\EOS\space}%
\providecommand \EOS [0]{\spacefactor3000\relax}%
\providecommand \BibitemShut  [1]{\csname bibitem#1\endcsname}%
\let\auto@bib@innerbib\@empty
\bibitem [{\citenamefont {Pohl}\ \emph {et~al.}(2002)\citenamefont {Pohl},
  \citenamefont {Liu},\ and\ \citenamefont {Thompson}}]{RevModPhys.74.991}%
  \BibitemOpen
  \bibfield  {author} {\bibinfo {author} {\bibfnamefont {R.~O.}\ \bibnamefont
  {Pohl}}, \bibinfo {author} {\bibfnamefont {X.}~\bibnamefont {Liu}},\ and\
  \bibinfo {author} {\bibfnamefont {E.}~\bibnamefont {Thompson}},\ }\href
  {https://doi.org/10.1103/RevModPhys.74.991} {\bibfield  {journal} {\bibinfo
  {journal} {Reviews of Modern Physics}\ }\textbf {\bibinfo {volume} {74}},\
  \bibinfo {pages} {991} (\bibinfo {year} {2002})}\BibitemShut {NoStop}%
\bibitem [{\citenamefont {Cahill}\ and\ \citenamefont
  {Pohl}(1987)}]{Cahill1987}%
  \BibitemOpen
  \bibfield  {author} {\bibinfo {author} {\bibfnamefont {D.~G.}\ \bibnamefont
  {Cahill}}\ and\ \bibinfo {author} {\bibfnamefont {R.~O.}\ \bibnamefont
  {Pohl}},\ }\href {https://doi.org/10.1103/PhysRevB.35.4067} {\bibfield
  {journal} {\bibinfo  {journal} {Physical Review B}\ }\textbf {\bibinfo
  {volume} {35}},\ \bibinfo {pages} {4067} (\bibinfo {year}
  {1987})}\BibitemShut {NoStop}%
\bibitem [{\citenamefont {Wingert}\ \emph {et~al.}(2016)\citenamefont
  {Wingert}, \citenamefont {Zheng}, \citenamefont {Kwon},\ and\ \citenamefont
  {Chen}}]{Wingert2016}%
  \BibitemOpen
  \bibfield  {author} {\bibinfo {author} {\bibfnamefont {M.~C.}\ \bibnamefont
  {Wingert}}, \bibinfo {author} {\bibfnamefont {J.}~\bibnamefont {Zheng}},
  \bibinfo {author} {\bibfnamefont {S.}~\bibnamefont {Kwon}},\ and\ \bibinfo
  {author} {\bibfnamefont {R.}~\bibnamefont {Chen}},\ }\href
  {https://doi.org/10.1088/0268-1242/31/11/113003} {\bibfield  {journal}
  {\bibinfo  {journal} {Semiconductor Science and Technology}\ }\textbf
  {\bibinfo {volume} {31}},\ \bibinfo {pages} {113003} (\bibinfo {year}
  {2016})}\BibitemShut {NoStop}%
\bibitem [{\citenamefont {DeAngelis}\ \emph {et~al.}(2019)\citenamefont
  {DeAngelis}, \citenamefont {Muraleedharan}, \citenamefont {Moon},
  \citenamefont {Seyf}, \citenamefont {Minnich}, \citenamefont {McGaughey},\
  and\ \citenamefont {Henry}}]{DeAngelis2019}%
  \BibitemOpen
  \bibfield  {author} {\bibinfo {author} {\bibfnamefont {F.}~\bibnamefont
  {DeAngelis}}, \bibinfo {author} {\bibfnamefont {M.~G.}\ \bibnamefont
  {Muraleedharan}}, \bibinfo {author} {\bibfnamefont {J.}~\bibnamefont {Moon}},
  \bibinfo {author} {\bibfnamefont {H.~R.}\ \bibnamefont {Seyf}}, \bibinfo
  {author} {\bibfnamefont {A.~J.}\ \bibnamefont {Minnich}}, \bibinfo {author}
  {\bibfnamefont {A.~J.}\ \bibnamefont {McGaughey}},\ and\ \bibinfo {author}
  {\bibfnamefont {A.}~\bibnamefont {Henry}},\ }\href
  {https://doi.org/10.1080/15567265.2018.1519004} {\bibfield  {journal}
  {\bibinfo  {journal} {Nanoscale and Microscale Thermophysical Engineering}\
  }\textbf {\bibinfo {volume} {23}},\ \bibinfo {pages} {81} (\bibinfo {year}
  {2019})}\BibitemShut {NoStop}%
\bibitem [{\citenamefont {Zhou}\ \emph {et~al.}(2020)\citenamefont {Zhou},
  \citenamefont {Cheng}, \citenamefont {Chen}, \citenamefont {Xie},
  \citenamefont {Wang},\ and\ \citenamefont {Zhang}}]{Zhou2020}%
  \BibitemOpen
  \bibfield  {author} {\bibinfo {author} {\bibfnamefont {W.-X.}\ \bibnamefont
  {Zhou}}, \bibinfo {author} {\bibfnamefont {Y.}~\bibnamefont {Cheng}},
  \bibinfo {author} {\bibfnamefont {K.-Q.}\ \bibnamefont {Chen}}, \bibinfo
  {author} {\bibfnamefont {G.}~\bibnamefont {Xie}}, \bibinfo {author}
  {\bibfnamefont {T.}~\bibnamefont {Wang}},\ and\ \bibinfo {author}
  {\bibfnamefont {G.}~\bibnamefont {Zhang}},\ }\href
  {https://doi.org/https://doi.org/10.1002/adfm.201903829} {\bibfield
  {journal} {\bibinfo  {journal} {Advanced Functional Materials}\ }\textbf
  {\bibinfo {volume} {30}},\ \bibinfo {pages} {1903829} (\bibinfo {year}
  {2020})}\BibitemShut {NoStop}%
\bibitem [{\citenamefont {Zhang}\ \emph {et~al.}(2020)\citenamefont {Zhang},
  \citenamefont {Ouyang}, \citenamefont {Cheng}, \citenamefont {Chen},
  \citenamefont {Li},\ and\ \citenamefont {Zhang}}]{Zhang2020}%
  \BibitemOpen
  \bibfield  {author} {\bibinfo {author} {\bibfnamefont {Z.}~\bibnamefont
  {Zhang}}, \bibinfo {author} {\bibfnamefont {Y.}~\bibnamefont {Ouyang}},
  \bibinfo {author} {\bibfnamefont {Y.}~\bibnamefont {Cheng}}, \bibinfo
  {author} {\bibfnamefont {J.}~\bibnamefont {Chen}}, \bibinfo {author}
  {\bibfnamefont {N.}~\bibnamefont {Li}},\ and\ \bibinfo {author}
  {\bibfnamefont {G.}~\bibnamefont {Zhang}},\ }\href
  {https://doi.org/10.1016/j.physrep.2020.03.001} {\bibfield  {journal}
  {\bibinfo  {journal} {Physics Reports}\ }\textbf {\bibinfo {volume} {860}},\
  \bibinfo {pages} {1} (\bibinfo {year} {2020})}\BibitemShut {NoStop}%
\bibitem [{\citenamefont {Feldman}\ \emph {et~al.}(1993)\citenamefont
  {Feldman}, \citenamefont {Kluge}, \citenamefont {Allen},\ and\ \citenamefont
  {Wooten}}]{Allen1993}%
  \BibitemOpen
  \bibfield  {author} {\bibinfo {author} {\bibfnamefont {J.~L.}\ \bibnamefont
  {Feldman}}, \bibinfo {author} {\bibfnamefont {M.~D.}\ \bibnamefont {Kluge}},
  \bibinfo {author} {\bibfnamefont {P.~B.}\ \bibnamefont {Allen}},\ and\
  \bibinfo {author} {\bibfnamefont {F.}~\bibnamefont {Wooten}},\ }\href
  {https://doi.org/10.1103/PhysRevB.48.12589} {\bibfield  {journal} {\bibinfo
  {journal} {Physical Review B}\ }\textbf {\bibinfo {volume} {48}},\ \bibinfo
  {pages} {12589} (\bibinfo {year} {1993})}\BibitemShut {NoStop}%
\bibitem [{\citenamefont {Allen}\ and\ \citenamefont
  {Feldman}(1993)}]{Allen1993a}%
  \BibitemOpen
  \bibfield  {author} {\bibinfo {author} {\bibfnamefont {P.~B.}\ \bibnamefont
  {Allen}}\ and\ \bibinfo {author} {\bibfnamefont {J.~L.}\ \bibnamefont
  {Feldman}},\ }\href {https://doi.org/10.1103/PhysRevB.48.12581} {\bibfield
  {journal} {\bibinfo  {journal} {Physical Review B}\ }\textbf {\bibinfo
  {volume} {48}},\ \bibinfo {pages} {12581} (\bibinfo {year}
  {1993})}\BibitemShut {NoStop}%
\bibitem [{\citenamefont {Allen}\ \emph {et~al.}(1999)\citenamefont {Allen},
  \citenamefont {Feldman}, \citenamefont {Fabian},\ and\ \citenamefont
  {Wooten}}]{Allen1999a}%
  \BibitemOpen
  \bibfield  {author} {\bibinfo {author} {\bibfnamefont {P.~B.}\ \bibnamefont
  {Allen}}, \bibinfo {author} {\bibfnamefont {J.~L.}\ \bibnamefont {Feldman}},
  \bibinfo {author} {\bibfnamefont {J.}~\bibnamefont {Fabian}},\ and\ \bibinfo
  {author} {\bibfnamefont {F.}~\bibnamefont {Wooten}},\ }\href
  {https://doi.org/10.1080/13642819908223054} {\bibfield  {journal} {\bibinfo
  {journal} {Philosophical Magazine B: Physics of Condensed Matter; Statistical
  Mechanics, Electronic, Optical and Magnetic Properties}\ }\textbf {\bibinfo
  {volume} {79}},\ \bibinfo {pages} {1715} (\bibinfo {year}
  {1999})}\BibitemShut {NoStop}%
\bibitem [{\citenamefont {Lv}\ and\ \citenamefont
  {Henry}(2016{\natexlab{a}})}]{Lv2016}%
  \BibitemOpen
  \bibfield  {author} {\bibinfo {author} {\bibfnamefont {W.}~\bibnamefont
  {Lv}}\ and\ \bibinfo {author} {\bibfnamefont {A.}~\bibnamefont {Henry}},\
  }\href {https://doi.org/10.1038/srep37675} {\bibfield  {journal} {\bibinfo
  {journal} {Scientific Reports}\ }\textbf {\bibinfo {volume} {6}},\ \bibinfo
  {pages} {37675} (\bibinfo {year} {2016}{\natexlab{a}})}\BibitemShut {NoStop}%
\bibitem [{\citenamefont {Lv}\ and\ \citenamefont
  {Henry}(2016{\natexlab{b}})}]{Lv2016-2}%
  \BibitemOpen
  \bibfield  {author} {\bibinfo {author} {\bibfnamefont {W.}~\bibnamefont
  {Lv}}\ and\ \bibinfo {author} {\bibfnamefont {A.}~\bibnamefont {Henry}},\
  }\href {https://doi.org/10.1038/srep35720} {\bibfield  {journal} {\bibinfo
  {journal} {Scientific Reports}\ }\textbf {\bibinfo {volume} {6}},\ \bibinfo
  {pages} {35720} (\bibinfo {year} {2016}{\natexlab{b}})}\BibitemShut {NoStop}%
\bibitem [{\citenamefont {Simoncelli}\ \emph {et~al.}(2019)\citenamefont
  {Simoncelli}, \citenamefont {Marzari},\ and\ \citenamefont
  {Mauri}}]{Simoncelli2019}%
  \BibitemOpen
  \bibfield  {author} {\bibinfo {author} {\bibfnamefont {M.}~\bibnamefont
  {Simoncelli}}, \bibinfo {author} {\bibfnamefont {N.}~\bibnamefont
  {Marzari}},\ and\ \bibinfo {author} {\bibfnamefont {F.}~\bibnamefont
  {Mauri}},\ }\href {https://doi.org/10.1038/s41567-019-0520-x} {\bibfield
  {journal} {\bibinfo  {journal} {Nature Physics}\ }\textbf {\bibinfo {volume}
  {15}},\ \bibinfo {pages} {809} (\bibinfo {year} {2019})}\BibitemShut
  {NoStop}%
\bibitem [{\citenamefont {Isaeva}\ \emph {et~al.}(2019)\citenamefont {Isaeva},
  \citenamefont {Barbalinardo}, \citenamefont {Donadio},\ and\ \citenamefont
  {Baroni}}]{Isaeva2019a}%
  \BibitemOpen
  \bibfield  {author} {\bibinfo {author} {\bibfnamefont {L.}~\bibnamefont
  {Isaeva}}, \bibinfo {author} {\bibfnamefont {G.}~\bibnamefont
  {Barbalinardo}}, \bibinfo {author} {\bibfnamefont {D.}~\bibnamefont
  {Donadio}},\ and\ \bibinfo {author} {\bibfnamefont {S.}~\bibnamefont
  {Baroni}},\ }\href {https://doi.org/10.1038/s41467-019-11572-4} {\bibfield
  {journal} {\bibinfo  {journal} {Nature Communications}\ }\textbf {\bibinfo
  {volume} {10}},\ \bibinfo {pages} {3853} (\bibinfo {year}
  {2019})}\BibitemShut {NoStop}%
\bibitem [{\citenamefont {Xi}\ \emph {et~al.}(2017)\citenamefont {Xi},
  \citenamefont {Zhang}, \citenamefont {Chen}, \citenamefont {Zhou},
  \citenamefont {Nakayama},\ and\ \citenamefont {Li}}]{Xi2017}%
  \BibitemOpen
  \bibfield  {author} {\bibinfo {author} {\bibfnamefont {Q.}~\bibnamefont
  {Xi}}, \bibinfo {author} {\bibfnamefont {Z.}~\bibnamefont {Zhang}}, \bibinfo
  {author} {\bibfnamefont {J.}~\bibnamefont {Chen}}, \bibinfo {author}
  {\bibfnamefont {J.}~\bibnamefont {Zhou}}, \bibinfo {author} {\bibfnamefont
  {T.}~\bibnamefont {Nakayama}},\ and\ \bibinfo {author} {\bibfnamefont
  {B.}~\bibnamefont {Li}},\ }\href {https://doi.org/10.1103/PhysRevB.96.064306}
  {\bibfield  {journal} {\bibinfo  {journal} {Physical Review B}\ }\textbf
  {\bibinfo {volume} {96}},\ \bibinfo {pages} {064306} (\bibinfo {year}
  {2017})}\BibitemShut {NoStop}%
\bibitem [{\citenamefont {Xi}\ \emph {et~al.}(2018)\citenamefont {Xi},
  \citenamefont {Zhang}, \citenamefont {Nakayama}, \citenamefont {Chen},
  \citenamefont {Zhou},\ and\ \citenamefont {Li}}]{Zhou2018a}%
  \BibitemOpen
  \bibfield  {author} {\bibinfo {author} {\bibfnamefont {Q.}~\bibnamefont
  {Xi}}, \bibinfo {author} {\bibfnamefont {Z.}~\bibnamefont {Zhang}}, \bibinfo
  {author} {\bibfnamefont {T.}~\bibnamefont {Nakayama}}, \bibinfo {author}
  {\bibfnamefont {J.}~\bibnamefont {Chen}}, \bibinfo {author} {\bibfnamefont
  {J.}~\bibnamefont {Zhou}},\ and\ \bibinfo {author} {\bibfnamefont
  {B.}~\bibnamefont {Li}},\ }\href {https://doi.org/10.1103/PhysRevB.97.224308}
  {\bibfield  {journal} {\bibinfo  {journal} {Physical Review B}\ }\textbf
  {\bibinfo {volume} {97}},\ \bibinfo {pages} {2} (\bibinfo {year}
  {2018})}\BibitemShut {NoStop}%
\bibitem [{\citenamefont {Bickham}(1999)}]{Bickham1999}%
  \BibitemOpen
  \bibfield  {author} {\bibinfo {author} {\bibfnamefont {S.}~\bibnamefont
  {Bickham}},\ }\href {https://doi.org/10.1103/PhysRevB.59.3551} {\bibfield
  {journal} {\bibinfo  {journal} {Physical Review B}\ }\textbf {\bibinfo
  {volume} {59}},\ \bibinfo {pages} {3551} (\bibinfo {year}
  {1999})}\BibitemShut {NoStop}%
\bibitem [{\citenamefont {{R.J. Hardy}}(1968)}]{R.J.Hardy1968}%
  \BibitemOpen
  \bibfield  {author} {\bibinfo {author} {\bibnamefont {{R.J. Hardy}}},\
  }\href@noop {} {\bibfield  {journal} {\bibinfo  {journal} {Physical Review}\
  }\textbf {\bibinfo {volume} {132}},\ \bibinfo {pages} {168} (\bibinfo {year}
  {1968})}\BibitemShut {NoStop}%
\bibitem [{\citenamefont {Shintani}\ and\ \citenamefont
  {Tanaka}(2008)}]{Shintani2008}%
  \BibitemOpen
  \bibfield  {author} {\bibinfo {author} {\bibfnamefont {H.}~\bibnamefont
  {Shintani}}\ and\ \bibinfo {author} {\bibfnamefont {H.}~\bibnamefont
  {Tanaka}},\ }\href {https://doi.org/10.1038/nmat2293} {\bibfield  {journal}
  {\bibinfo  {journal} {Nature Materials}\ }\textbf {\bibinfo {volume} {7}},\
  \bibinfo {pages} {870} (\bibinfo {year} {2008})}\BibitemShut {NoStop}%
\bibitem [{\citenamefont {Larkin}\ and\ \citenamefont
  {McGaughey}(2014)}]{Larkin2014}%
  \BibitemOpen
  \bibfield  {author} {\bibinfo {author} {\bibfnamefont {J.~M.}\ \bibnamefont
  {Larkin}}\ and\ \bibinfo {author} {\bibfnamefont {A.~J.}\ \bibnamefont
  {McGaughey}},\ }\href {https://doi.org/10.1103/PhysRevB.89.144303} {\bibfield
   {journal} {\bibinfo  {journal} {Physical Review B}\ }\textbf {\bibinfo
  {volume} {89}},\ \bibinfo {pages} {144303} (\bibinfo {year}
  {2014})}\BibitemShut {NoStop}%
\bibitem [{\citenamefont {Moon}\ \emph {et~al.}(2018)\citenamefont {Moon},
  \citenamefont {Latour},\ and\ \citenamefont {Minnich}}]{Moon2018b}%
  \BibitemOpen
  \bibfield  {author} {\bibinfo {author} {\bibfnamefont {J.}~\bibnamefont
  {Moon}}, \bibinfo {author} {\bibfnamefont {B.}~\bibnamefont {Latour}},\ and\
  \bibinfo {author} {\bibfnamefont {A.~J.}\ \bibnamefont {Minnich}},\ }\href
  {https://doi.org/10.1103/PhysRevB.97.024201} {\bibfield  {journal} {\bibinfo
  {journal} {Physical Review B}\ }\textbf {\bibinfo {volume} {97}},\ \bibinfo
  {pages} {024201} (\bibinfo {year} {2018})}\BibitemShut {NoStop}%
\bibitem [{\citenamefont {Moon}\ \emph {et~al.}(2019)\citenamefont {Moon},
  \citenamefont {Hermann}, \citenamefont {Manley}, \citenamefont {Alatas},
  \citenamefont {Said},\ and\ \citenamefont {Minnich}}]{Moon2019}%
  \BibitemOpen
  \bibfield  {author} {\bibinfo {author} {\bibfnamefont {J.}~\bibnamefont
  {Moon}}, \bibinfo {author} {\bibfnamefont {R.~P.}\ \bibnamefont {Hermann}},
  \bibinfo {author} {\bibfnamefont {M.~E.}\ \bibnamefont {Manley}}, \bibinfo
  {author} {\bibfnamefont {A.}~\bibnamefont {Alatas}}, \bibinfo {author}
  {\bibfnamefont {A.~H.}\ \bibnamefont {Said}},\ and\ \bibinfo {author}
  {\bibfnamefont {A.~J.}\ \bibnamefont {Minnich}},\ }\href
  {https://doi.org/10.1103/PhysRevMaterials.3.065601} {\bibfield  {journal}
  {\bibinfo  {journal} {Physical Review Materials}\ }\textbf {\bibinfo {volume}
  {3}},\ \bibinfo {pages} {065601} (\bibinfo {year} {2019})}\BibitemShut
  {NoStop}%
\bibitem [{\citenamefont {Larkin}\ \emph {et~al.}(2014)\citenamefont {Larkin},
  \citenamefont {Turney}, \citenamefont {Massicotte}, \citenamefont {Amon},\
  and\ \citenamefont {McGaughey}}]{Larklin2020}%
  \BibitemOpen
  \bibfield  {author} {\bibinfo {author} {\bibfnamefont {J.~M.}\ \bibnamefont
  {Larkin}}, \bibinfo {author} {\bibfnamefont {J.~E.}\ \bibnamefont {Turney}},
  \bibinfo {author} {\bibfnamefont {A.~D.}\ \bibnamefont {Massicotte}},
  \bibinfo {author} {\bibfnamefont {C.~H.}\ \bibnamefont {Amon}},\ and\
  \bibinfo {author} {\bibfnamefont {A.~J.}\ \bibnamefont {McGaughey}},\ }\href
  {https://doi.org/10.1166/jctn.2014.3345} {\bibfield  {journal} {\bibinfo
  {journal} {Journal of Computational and Theoretical Nanoscience}\ }\textbf
  {\bibinfo {volume} {11}},\ \bibinfo {pages} {249} (\bibinfo {year}
  {2014})}\BibitemShut {NoStop}%
\bibitem [{\citenamefont {Kim}\ \emph {et~al.}(2021)\citenamefont {Kim},
  \citenamefont {Moon},\ and\ \citenamefont {Minnich}}]{Kim2021}%
  \BibitemOpen
  \bibfield  {author} {\bibinfo {author} {\bibfnamefont {T.}~\bibnamefont
  {Kim}}, \bibinfo {author} {\bibfnamefont {J.}~\bibnamefont {Moon}},\ and\
  \bibinfo {author} {\bibfnamefont {A.~J.}\ \bibnamefont {Minnich}},\ }\href
  {https://doi.org/10.1103/PhysRevMaterials.5.065602} {\bibfield  {journal}
  {\bibinfo  {journal} {Physical Review Materials}\ }\textbf {\bibinfo {volume}
  {5}},\ \bibinfo {pages} {65602} (\bibinfo {year} {2021})}\BibitemShut
  {NoStop}%
\bibitem [{\citenamefont {Shenogin}\ \emph {et~al.}(2009)\citenamefont
  {Shenogin}, \citenamefont {Bodapati}, \citenamefont {Keblinski},\ and\
  \citenamefont {McGaughey}}]{Shenogin2009}%
  \BibitemOpen
  \bibfield  {author} {\bibinfo {author} {\bibfnamefont {S.}~\bibnamefont
  {Shenogin}}, \bibinfo {author} {\bibfnamefont {A.}~\bibnamefont {Bodapati}},
  \bibinfo {author} {\bibfnamefont {P.}~\bibnamefont {Keblinski}},\ and\
  \bibinfo {author} {\bibfnamefont {A.~J.}\ \bibnamefont {McGaughey}},\ }\href
  {https://doi.org/10.1063/1.3073954} {\bibfield  {journal} {\bibinfo
  {journal} {Journal of Applied Physics}\ }\textbf {\bibinfo {volume} {105}},\
  \bibinfo {pages} {034906} (\bibinfo {year} {2009})}\BibitemShut {NoStop}%
\bibitem [{\citenamefont {Park}\ \emph {et~al.}(2014)\citenamefont {Park},
  \citenamefont {Lee},\ and\ \citenamefont {Kim}}]{Park2014}%
  \BibitemOpen
  \bibfield  {author} {\bibinfo {author} {\bibfnamefont {M.}~\bibnamefont
  {Park}}, \bibinfo {author} {\bibfnamefont {I.~H.}\ \bibnamefont {Lee}},\ and\
  \bibinfo {author} {\bibfnamefont {Y.~S.}\ \bibnamefont {Kim}},\ }\href
  {https://doi.org/10.1063/1.4891500} {\bibfield  {journal} {\bibinfo
  {journal} {Journal of Applied Physics}\ }\textbf {\bibinfo {volume} {116}},\
  \bibinfo {pages} {043514} (\bibinfo {year} {2014})}\BibitemShut {NoStop}%
\bibitem [{\citenamefont {S{\"{a}}{\"{a}}skilahti}\ \emph
  {et~al.}(2016)\citenamefont {S{\"{a}}{\"{a}}skilahti}, \citenamefont
  {Oksanen}, \citenamefont {Tulkki}, \citenamefont {McGaughey},\ and\
  \citenamefont {Volz}}]{Saaskilahti2016a}%
  \BibitemOpen
  \bibfield  {author} {\bibinfo {author} {\bibfnamefont {K.}~\bibnamefont
  {S{\"{a}}{\"{a}}skilahti}}, \bibinfo {author} {\bibfnamefont
  {J.}~\bibnamefont {Oksanen}}, \bibinfo {author} {\bibfnamefont
  {J.}~\bibnamefont {Tulkki}}, \bibinfo {author} {\bibfnamefont {A.~J.}\
  \bibnamefont {McGaughey}},\ and\ \bibinfo {author} {\bibfnamefont
  {S.}~\bibnamefont {Volz}},\ }\href {https://doi.org/10.1063/1.4968617}
  {\bibfield  {journal} {\bibinfo  {journal} {AIP Advances}\ }\textbf {\bibinfo
  {volume} {6}},\ \bibinfo {pages} {121904} (\bibinfo {year}
  {2016})}\BibitemShut {NoStop}%
\bibitem [{\citenamefont {Zhou}(2021)}]{Zhou2021}%
  \BibitemOpen
  \bibfield  {author} {\bibinfo {author} {\bibfnamefont {Y.}~\bibnamefont
  {Zhou}},\ }\href {https://doi.org/10.1063/5.0054039} {\bibfield  {journal}
  {\bibinfo  {journal} {Journal of Applied Physics}\ }\textbf {\bibinfo
  {volume} {129}},\ \bibinfo {pages} {235104} (\bibinfo {year}
  {2021})}\BibitemShut {NoStop}%
\bibitem [{\citenamefont {Zhang}\ \emph
  {et~al.}(2021{\natexlab{a}})\citenamefont {Zhang}, \citenamefont {Guo},
  \citenamefont {Bescond}, \citenamefont {Chen}, \citenamefont {Nomura},\ and\
  \citenamefont {Volz}}]{zhang2021heat}%
  \BibitemOpen
  \bibfield  {author} {\bibinfo {author} {\bibfnamefont {Z.}~\bibnamefont
  {Zhang}}, \bibinfo {author} {\bibfnamefont {Y.}~\bibnamefont {Guo}}, \bibinfo
  {author} {\bibfnamefont {M.}~\bibnamefont {Bescond}}, \bibinfo {author}
  {\bibfnamefont {J.}~\bibnamefont {Chen}}, \bibinfo {author} {\bibfnamefont
  {M.}~\bibnamefont {Nomura}},\ and\ \bibinfo {author} {\bibfnamefont
  {S.}~\bibnamefont {Volz}},\ }\href@noop {} {\bibinfo {title} {Heat conduction
  theory including phonon coherence}} (\bibinfo {year} {2021}{\natexlab{a}}),\
  \Eprint {https://arxiv.org/abs/2110.09750} {arXiv:2110.09750
  [physics.comp-ph]} \BibitemShut {NoStop}%
\bibitem [{\citenamefont {Plimpton}(1995)}]{Plimpton1995}%
  \BibitemOpen
  \bibfield  {author} {\bibinfo {author} {\bibfnamefont {S.}~\bibnamefont
  {Plimpton}},\ }\href {https://doi.org/https://doi.org/10.1006/jcph.1995.1039}
  {\bibfield  {journal} {\bibinfo  {journal} {Journal of Computational
  Physics}\ }\textbf {\bibinfo {volume} {117}},\ \bibinfo {pages} {1} (\bibinfo
  {year} {1995})}\BibitemShut {NoStop}%
\bibitem [{\citenamefont {Stillinger}\ and\ \citenamefont
  {Weber}(1985)}]{Stillinger1985}%
  \BibitemOpen
  \bibfield  {author} {\bibinfo {author} {\bibfnamefont {F.~H.}\ \bibnamefont
  {Stillinger}}\ and\ \bibinfo {author} {\bibfnamefont {T.~A.}\ \bibnamefont
  {Weber}},\ }\href {https://doi.org/10.1103/PhysRevB.31.5262} {\bibfield
  {journal} {\bibinfo  {journal} {Physical Review B}\ }\textbf {\bibinfo
  {volume} {31}},\ \bibinfo {pages} {5262} (\bibinfo {year}
  {1985})}\BibitemShut {NoStop}%
\bibitem [{\citenamefont {France-Lanord}\ \emph {et~al.}(2014)\citenamefont
  {France-Lanord}, \citenamefont {Blandre}, \citenamefont {Albaret},
  \citenamefont {Merabia}, \citenamefont {Lacroix},\ and\ \citenamefont
  {Termentzidis}}]{France_Lanord_2014}%
  \BibitemOpen
  \bibfield  {author} {\bibinfo {author} {\bibfnamefont {A.}~\bibnamefont
  {France-Lanord}}, \bibinfo {author} {\bibfnamefont {E.}~\bibnamefont
  {Blandre}}, \bibinfo {author} {\bibfnamefont {T.}~\bibnamefont {Albaret}},
  \bibinfo {author} {\bibfnamefont {S.}~\bibnamefont {Merabia}}, \bibinfo
  {author} {\bibfnamefont {D.}~\bibnamefont {Lacroix}},\ and\ \bibinfo {author}
  {\bibfnamefont {K.}~\bibnamefont {Termentzidis}},\ }\href
  {https://doi.org/10.1088/0953-8984/26/5/055011} {\bibfield  {journal}
  {\bibinfo  {journal} {Journal of Physics: Condensed Matter}\ }\textbf
  {\bibinfo {volume} {26}},\ \bibinfo {pages} {55011} (\bibinfo {year}
  {2014})}\BibitemShut {NoStop}%
\bibitem [{\citenamefont {Boon}\ and\ \citenamefont {Yip}(1991)}]{boon1991}%
  \BibitemOpen
  \bibfield  {author} {\bibinfo {author} {\bibfnamefont {J.~P.}\ \bibnamefont
  {Boon}}\ and\ \bibinfo {author} {\bibfnamefont {S.}~\bibnamefont {Yip}},\
  }\href@noop {} {\emph {\bibinfo {title} {Molecular hydrodynamics}}}\
  (\bibinfo  {publisher} {Courier Corporation},\ \bibinfo {year}
  {1991})\BibitemShut {NoStop}%
\bibitem [{\citenamefont {Kubo}(1957)}]{Kubo1957}%
  \BibitemOpen
  \bibfield  {author} {\bibinfo {author} {\bibfnamefont {R.}~\bibnamefont
  {Kubo}},\ }\href {https://doi.org/10.1143/JPSJ.12.570} {\bibfield  {journal}
  {\bibinfo  {journal} {Journal of the Physical Society of Japan}\ }\textbf
  {\bibinfo {volume} {12}},\ \bibinfo {pages} {570} (\bibinfo {year}
  {1957})}\BibitemShut {NoStop}%
\bibitem [{\citenamefont {Kubo}\ \emph {et~al.}(1957)\citenamefont {Kubo},
  \citenamefont {Yokota},\ and\ \citenamefont {Nakajima}}]{Kubo1957a}%
  \BibitemOpen
  \bibfield  {author} {\bibinfo {author} {\bibfnamefont {R.}~\bibnamefont
  {Kubo}}, \bibinfo {author} {\bibfnamefont {M.}~\bibnamefont {Yokota}},\ and\
  \bibinfo {author} {\bibfnamefont {S.}~\bibnamefont {Nakajima}},\ }\href
  {https://doi.org/10.1143/JPSJ.12.1203} {\bibfield  {journal} {\bibinfo
  {journal} {Journal of the Physical Society of Japan}\ }\textbf {\bibinfo
  {volume} {12}},\ \bibinfo {pages} {1203} (\bibinfo {year}
  {1957})}\BibitemShut {NoStop}%
\bibitem [{\citenamefont {Volz}\ and\ \citenamefont {Chen}(2000)}]{Volz2000-1}%
  \BibitemOpen
  \bibfield  {author} {\bibinfo {author} {\bibfnamefont {S.~G.}\ \bibnamefont
  {Volz}}\ and\ \bibinfo {author} {\bibfnamefont {G.}~\bibnamefont {Chen}},\
  }\href {https://doi.org/10.1103/PhysRevB.61.2651} {\bibfield  {journal}
  {\bibinfo  {journal} {Physical Review B}\ }\textbf {\bibinfo {volume} {61}},\
  \bibinfo {pages} {2651} (\bibinfo {year} {2000})}\BibitemShut {NoStop}%
\bibitem [{\citenamefont {Torii}\ \emph {et~al.}(2008)\citenamefont {Torii},
  \citenamefont {Nakano},\ and\ \citenamefont {Ohara}}]{Torii2008}%
  \BibitemOpen
  \bibfield  {author} {\bibinfo {author} {\bibfnamefont {D.}~\bibnamefont
  {Torii}}, \bibinfo {author} {\bibfnamefont {T.}~\bibnamefont {Nakano}},\ and\
  \bibinfo {author} {\bibfnamefont {T.}~\bibnamefont {Ohara}},\ }\href
  {https://doi.org/10.1063/1.2821963} {\bibfield  {journal} {\bibinfo
  {journal} {The Journal of Chemical Physics}\ }\textbf {\bibinfo {volume}
  {128}},\ \bibinfo {pages} {44504} (\bibinfo {year} {2008})}\BibitemShut
  {NoStop}%
\bibitem [{\citenamefont {Schelling}\ \emph {et~al.}(2002)\citenamefont
  {Schelling}, \citenamefont {Phillpot},\ and\ \citenamefont
  {Keblinski}}]{Schelling2002}%
  \BibitemOpen
  \bibfield  {author} {\bibinfo {author} {\bibfnamefont {P.~K.}\ \bibnamefont
  {Schelling}}, \bibinfo {author} {\bibfnamefont {S.~R.}\ \bibnamefont
  {Phillpot}},\ and\ \bibinfo {author} {\bibfnamefont {P.}~\bibnamefont
  {Keblinski}},\ }\href {https://doi.org/10.1063/1.1465106} {\bibfield
  {journal} {\bibinfo  {journal} {Applied Physics Letters}\ }\textbf {\bibinfo
  {volume} {80}},\ \bibinfo {pages} {2484} (\bibinfo {year}
  {2002})}\BibitemShut {NoStop}%
\bibitem [{\citenamefont {Hu}\ \emph {et~al.}(2018)\citenamefont {Hu},
  \citenamefont {Zhang}, \citenamefont {Jiang}, \citenamefont {Chen},
  \citenamefont {Volz}, \citenamefont {Nomura},\ and\ \citenamefont
  {Li}}]{Hu2018}%
  \BibitemOpen
  \bibfield  {author} {\bibinfo {author} {\bibfnamefont {S.}~\bibnamefont
  {Hu}}, \bibinfo {author} {\bibfnamefont {Z.}~\bibnamefont {Zhang}}, \bibinfo
  {author} {\bibfnamefont {P.}~\bibnamefont {Jiang}}, \bibinfo {author}
  {\bibfnamefont {J.}~\bibnamefont {Chen}}, \bibinfo {author} {\bibfnamefont
  {S.}~\bibnamefont {Volz}}, \bibinfo {author} {\bibfnamefont {M.}~\bibnamefont
  {Nomura}},\ and\ \bibinfo {author} {\bibfnamefont {B.}~\bibnamefont {Li}},\
  }\href {https://doi.org/10.1021/acs.jpclett.8b01653} {\bibfield  {journal}
  {\bibinfo  {journal} {Journal of Physical Chemistry Letters}\ }\textbf
  {\bibinfo {volume} {9}},\ \bibinfo {pages} {3959} (\bibinfo {year}
  {2018})}\BibitemShut {NoStop}%
\bibitem [{\citenamefont {Jiang}\ \emph {et~al.}(2021)\citenamefont {Jiang},
  \citenamefont {Ouyang}, \citenamefont {Ren}, \citenamefont {Yu},
  \citenamefont {He},\ and\ \citenamefont {Chen}}]{Jiang2021}%
  \BibitemOpen
  \bibfield  {author} {\bibinfo {author} {\bibfnamefont {P.}~\bibnamefont
  {Jiang}}, \bibinfo {author} {\bibfnamefont {Y.}~\bibnamefont {Ouyang}},
  \bibinfo {author} {\bibfnamefont {W.}~\bibnamefont {Ren}}, \bibinfo {author}
  {\bibfnamefont {C.}~\bibnamefont {Yu}}, \bibinfo {author} {\bibfnamefont
  {J.}~\bibnamefont {He}},\ and\ \bibinfo {author} {\bibfnamefont
  {J.}~\bibnamefont {Chen}},\ }\href {https://doi.org/10.1063/5.0046509}
  {\bibfield  {journal} {\bibinfo  {journal} {APL Materials}\ }\textbf
  {\bibinfo {volume} {9}},\ \bibinfo {pages} {040703} (\bibinfo {year}
  {2021})}\BibitemShut {NoStop}%
\bibitem [{\citenamefont {Zhang}\ \emph
  {et~al.}(2021{\natexlab{b}})\citenamefont {Zhang}, \citenamefont {Guo},
  \citenamefont {Bescond}, \citenamefont {Chen}, \citenamefont {Nomura},\ and\
  \citenamefont {Volz}}]{Zhang2021}%
  \BibitemOpen
  \bibfield  {author} {\bibinfo {author} {\bibfnamefont {Z.}~\bibnamefont
  {Zhang}}, \bibinfo {author} {\bibfnamefont {Y.}~\bibnamefont {Guo}}, \bibinfo
  {author} {\bibfnamefont {M.}~\bibnamefont {Bescond}}, \bibinfo {author}
  {\bibfnamefont {J.}~\bibnamefont {Chen}}, \bibinfo {author} {\bibfnamefont
  {M.}~\bibnamefont {Nomura}},\ and\ \bibinfo {author} {\bibfnamefont
  {S.}~\bibnamefont {Volz}},\ }\href
  {https://doi.org/10.1103/PhysRevB.103.184307} {\bibfield  {journal} {\bibinfo
   {journal} {Physical Review B}\ }\textbf {\bibinfo {volume} {103}},\ \bibinfo
  {pages} {184307} (\bibinfo {year} {2021}{\natexlab{b}})}\BibitemShut
  {NoStop}%
\bibitem [{\citenamefont {Shiomi}\ and\ \citenamefont
  {Maruyama}(2006)}]{Shiomi2006}%
  \BibitemOpen
  \bibfield  {author} {\bibinfo {author} {\bibfnamefont {J.}~\bibnamefont
  {Shiomi}}\ and\ \bibinfo {author} {\bibfnamefont {S.}~\bibnamefont
  {Maruyama}},\ }\href {https://doi.org/10.1103/PhysRevB.73.205420} {\bibfield
  {journal} {\bibinfo  {journal} {Physical Review B}\ }\textbf {\bibinfo
  {volume} {73}},\ \bibinfo {pages} {205420} (\bibinfo {year}
  {2006})}\BibitemShut {NoStop}%
\bibitem [{\citenamefont {Baker}\ \emph {et~al.}(2012)\citenamefont {Baker},
  \citenamefont {Jordan},\ and\ \citenamefont {Norris}}]{Baker2012}%
  \BibitemOpen
  \bibfield  {author} {\bibinfo {author} {\bibfnamefont {C.~H.}\ \bibnamefont
  {Baker}}, \bibinfo {author} {\bibfnamefont {D.~A.}\ \bibnamefont {Jordan}},\
  and\ \bibinfo {author} {\bibfnamefont {P.~M.}\ \bibnamefont {Norris}},\
  }\href {https://doi.org/10.1103/PhysRevB.86.104306} {\bibfield  {journal}
  {\bibinfo  {journal} {Phys. Rev. B}\ }\textbf {\bibinfo {volume} {86}},\
  \bibinfo {pages} {104306} (\bibinfo {year} {2012})}\BibitemShut {NoStop}%
\bibitem [{\citenamefont {Moon}(2021)}]{moon2021examining}%
  \BibitemOpen
  \bibfield  {author} {\bibinfo {author} {\bibfnamefont {J.}~\bibnamefont
  {Moon}},\ }\href@noop {} {\bibfield  {journal} {\bibinfo  {journal} {arXiv
  preprint arXiv:2106.08459}\ } (\bibinfo {year} {2021})}\BibitemShut {NoStop}%
\bibitem [{\citenamefont {Beltukov}\ \emph {et~al.}(2013)\citenamefont
  {Beltukov}, \citenamefont {Kozub},\ and\ \citenamefont
  {Parshin}}]{Beltukov2013}%
  \BibitemOpen
  \bibfield  {author} {\bibinfo {author} {\bibfnamefont {Y.~M.}\ \bibnamefont
  {Beltukov}}, \bibinfo {author} {\bibfnamefont {V.~I.}\ \bibnamefont
  {Kozub}},\ and\ \bibinfo {author} {\bibfnamefont {D.~A.}\ \bibnamefont
  {Parshin}},\ }\href {https://doi.org/10.1103/PhysRevB.87.134203} {\bibfield
  {journal} {\bibinfo  {journal} {Physical Review B}\ }\textbf {\bibinfo
  {volume} {87}},\ \bibinfo {pages} {1} (\bibinfo {year} {2013})}\BibitemShut
  {NoStop}%
\bibitem [{\citenamefont {Zhu}\ and\ \citenamefont {Ertekin}(2016)}]{Zhu2016}%
  \BibitemOpen
  \bibfield  {author} {\bibinfo {author} {\bibfnamefont {T.}~\bibnamefont
  {Zhu}}\ and\ \bibinfo {author} {\bibfnamefont {E.}~\bibnamefont {Ertekin}},\
  }\href {https://doi.org/10.1021/acs.nanolett.6b00557} {\bibfield  {journal}
  {\bibinfo  {journal} {Nano Letters}\ }\textbf {\bibinfo {volume} {16}},\
  \bibinfo {pages} {4763} (\bibinfo {year} {2016})}\BibitemShut {NoStop}%
\bibitem [{\citenamefont {Seyf}\ and\ \citenamefont {Henry}(2016)}]{Seyf2016}%
  \BibitemOpen
  \bibfield  {author} {\bibinfo {author} {\bibfnamefont {H.~R.}\ \bibnamefont
  {Seyf}}\ and\ \bibinfo {author} {\bibfnamefont {A.}~\bibnamefont {Henry}},\
  }\href {https://doi.org/10.1063/1.4955420} {\bibfield  {journal} {\bibinfo
  {journal} {Journal of Applied Physics}\ }\textbf {\bibinfo {volume} {120}},\
  \bibinfo {pages} {025101} (\bibinfo {year} {2016})}\BibitemShut {NoStop}%
\bibitem [{\citenamefont {Zink}\ \emph {et~al.}(2006)\citenamefont {Zink},
  \citenamefont {Pietri},\ and\ \citenamefont {Hellman}}]{Zink2006}%
  \BibitemOpen
  \bibfield  {author} {\bibinfo {author} {\bibfnamefont {B.~L.}\ \bibnamefont
  {Zink}}, \bibinfo {author} {\bibfnamefont {R.}~\bibnamefont {Pietri}},\ and\
  \bibinfo {author} {\bibfnamefont {F.}~\bibnamefont {Hellman}},\ }\href
  {https://doi.org/10.1103/PhysRevLett.96.055902} {\bibfield  {journal}
  {\bibinfo  {journal} {Physical Review Letters}\ }\textbf {\bibinfo {volume}
  {96}},\ \bibinfo {pages} {055902} (\bibinfo {year} {2006})}\BibitemShut
  {NoStop}%
\bibitem [{\citenamefont {Gelin}\ \emph {et~al.}(2016)\citenamefont {Gelin},
  \citenamefont {Tanaka},\ and\ \citenamefont {Lema{\^{i}}tre}}]{Gelin2016}%
  \BibitemOpen
  \bibfield  {author} {\bibinfo {author} {\bibfnamefont {S.}~\bibnamefont
  {Gelin}}, \bibinfo {author} {\bibfnamefont {H.}~\bibnamefont {Tanaka}},\ and\
  \bibinfo {author} {\bibfnamefont {A.}~\bibnamefont {Lema{\^{i}}tre}},\ }\href
  {https://doi.org/10.1038/nmat4736} {\bibfield  {journal} {\bibinfo  {journal}
  {Nature Materials}\ }\textbf {\bibinfo {volume} {15}},\ \bibinfo {pages}
  {1177} (\bibinfo {year} {2016})}\BibitemShut {NoStop}%
\bibitem [{\citenamefont {Zhou}\ and\ \citenamefont {Hu}(2016)}]{Zhou2016b}%
  \BibitemOpen
  \bibfield  {author} {\bibinfo {author} {\bibfnamefont {Y.}~\bibnamefont
  {Zhou}}\ and\ \bibinfo {author} {\bibfnamefont {M.}~\bibnamefont {Hu}},\
  }\href {https://doi.org/10.1021/acs.nanolett.6b02450} {\bibfield  {journal}
  {\bibinfo  {journal} {Nano Letters}\ }\textbf {\bibinfo {volume} {16}},\
  \bibinfo {pages} {6178} (\bibinfo {year} {2016})}\BibitemShut {NoStop}%
\bibitem [{\citenamefont {Zhou}\ \emph {et~al.}(2017)\citenamefont {Zhou},
  \citenamefont {Morshedifard}, \citenamefont {Lee},\ and\ \citenamefont
  {{Abdolhosseini Qomi}}}]{Zhou2017b}%
  \BibitemOpen
  \bibfield  {author} {\bibinfo {author} {\bibfnamefont {Y.}~\bibnamefont
  {Zhou}}, \bibinfo {author} {\bibfnamefont {A.}~\bibnamefont {Morshedifard}},
  \bibinfo {author} {\bibfnamefont {J.}~\bibnamefont {Lee}},\ and\ \bibinfo
  {author} {\bibfnamefont {M.~J.}\ \bibnamefont {{Abdolhosseini Qomi}}},\
  }\href {https://doi.org/10.1063/1.4975159} {\bibfield  {journal} {\bibinfo
  {journal} {Applied Physics Letters}\ }\textbf {\bibinfo {volume} {110}},\
  \bibinfo {pages} {043104} (\bibinfo {year} {2017})}\BibitemShut {NoStop}%
\bibitem [{\citenamefont {Ziman}(2001)}]{ziman2001}%
  \BibitemOpen
  \bibfield  {author} {\bibinfo {author} {\bibfnamefont {J.~M.}\ \bibnamefont
  {Ziman}},\ }\href@noop {} {\emph {\bibinfo {title} {Electrons and phonons:
  the theory of transport phenomena in solids}}}\ (\bibinfo  {publisher}
  {Oxford university press},\ \bibinfo {year} {2001})\BibitemShut {NoStop}%
\bibitem [{\citenamefont {Gale}\ and\ \citenamefont {Rohl}(2003)}]{GULP}%
  \BibitemOpen
  \bibfield  {author} {\bibinfo {author} {\bibfnamefont {J.~D.}\ \bibnamefont
  {Gale}}\ and\ \bibinfo {author} {\bibfnamefont {A.~L.}\ \bibnamefont
  {Rohl}},\ }\href {https://doi.org/10.1080/0892702031000104887} {\bibfield
  {journal} {\bibinfo  {journal} {Molecular Simulation}\ }\textbf {\bibinfo
  {volume} {29}},\ \bibinfo {pages} {291} (\bibinfo {year} {2003})}\BibitemShut
  {NoStop}%
\bibitem [{\citenamefont {Barbalinardo}\ \emph {et~al.}(2020)\citenamefont
  {Barbalinardo}, \citenamefont {Chen}, \citenamefont {Lundgren},\ and\
  \citenamefont {Donadio}}]{kALDo}%
  \BibitemOpen
  \bibfield  {author} {\bibinfo {author} {\bibfnamefont {G.}~\bibnamefont
  {Barbalinardo}}, \bibinfo {author} {\bibfnamefont {Z.}~\bibnamefont {Chen}},
  \bibinfo {author} {\bibfnamefont {N.~W.}\ \bibnamefont {Lundgren}},\ and\
  \bibinfo {author} {\bibfnamefont {D.}~\bibnamefont {Donadio}},\ }\href
  {https://doi.org/10.1063/5.0020443} {\bibfield  {journal} {\bibinfo
  {journal} {Journal of Applied Physics}\ }\textbf {\bibinfo {volume} {128}},\
  \bibinfo {pages} {135104} (\bibinfo {year} {2020})}\BibitemShut {NoStop}%
\bibitem [{\citenamefont {Stokes}(2010)}]{Casimir1938}%
  \BibitemOpen
  \bibfield  {author} {\bibinfo {author} {\bibfnamefont {G.~G.}\ \bibnamefont
  {Stokes}},\ }\href {https://doi.org/10.1017/cbo9780511702266.008} {\bibfield
  {journal} {\bibinfo  {journal} {Mathematical and Physical Papers}\ }\textbf
  {\bibinfo {volume} {5}},\ \bibinfo {pages} {203} (\bibinfo {year}
  {2010})}\BibitemShut {NoStop}%
\bibitem [{\citenamefont {Zhang}\ \emph {et~al.}(2017)\citenamefont {Zhang},
  \citenamefont {Hu}, \citenamefont {Chen},\ and\ \citenamefont
  {Li}}]{Zhang2017k}%
  \BibitemOpen
  \bibfield  {author} {\bibinfo {author} {\bibfnamefont {Z.}~\bibnamefont
  {Zhang}}, \bibinfo {author} {\bibfnamefont {S.}~\bibnamefont {Hu}}, \bibinfo
  {author} {\bibfnamefont {J.}~\bibnamefont {Chen}},\ and\ \bibinfo {author}
  {\bibfnamefont {B.}~\bibnamefont {Li}},\ }\href
  {https://doi.org/10.1088/1361-6528/aa6e49} {\bibfield  {journal} {\bibinfo
  {journal} {Nanotechnology}\ }\textbf {\bibinfo {volume} {28}},\ \bibinfo
  {pages} {225704} (\bibinfo {year} {2017})}\BibitemShut {NoStop}%
\bibitem [{\citenamefont {Regner}\ \emph {et~al.}(2013)\citenamefont {Regner},
  \citenamefont {Sellan}, \citenamefont {Su}, \citenamefont {Amon},
  \citenamefont {McGaughey},\ and\ \citenamefont {Malen}}]{Regner2013}%
  \BibitemOpen
  \bibfield  {author} {\bibinfo {author} {\bibfnamefont {K.~T.}\ \bibnamefont
  {Regner}}, \bibinfo {author} {\bibfnamefont {D.~P.}\ \bibnamefont {Sellan}},
  \bibinfo {author} {\bibfnamefont {Z.}~\bibnamefont {Su}}, \bibinfo {author}
  {\bibfnamefont {C.~H.}\ \bibnamefont {Amon}}, \bibinfo {author}
  {\bibfnamefont {A.~J.}\ \bibnamefont {McGaughey}},\ and\ \bibinfo {author}
  {\bibfnamefont {J.~A.}\ \bibnamefont {Malen}},\ }\href
  {https://doi.org/10.1038/ncomms2630} {\bibfield  {journal} {\bibinfo
  {journal} {Nature Communications}\ }\textbf {\bibinfo {volume} {4}},\
  \bibinfo {pages} {1640} (\bibinfo {year} {2013})}\BibitemShut {NoStop}%
\bibitem [{\citenamefont {Braun}\ \emph {et~al.}(2016)\citenamefont {Braun},
  \citenamefont {Baker}, \citenamefont {Giri}, \citenamefont {Elahi},
  \citenamefont {Artyushkova}, \citenamefont {Beechem}, \citenamefont {Norris},
  \citenamefont {Leseman}, \citenamefont {Gaskins},\ and\ \citenamefont
  {Hopkins}}]{Braun2016}%
  \BibitemOpen
  \bibfield  {author} {\bibinfo {author} {\bibfnamefont {J.~L.}\ \bibnamefont
  {Braun}}, \bibinfo {author} {\bibfnamefont {C.~H.}\ \bibnamefont {Baker}},
  \bibinfo {author} {\bibfnamefont {A.}~\bibnamefont {Giri}}, \bibinfo {author}
  {\bibfnamefont {M.}~\bibnamefont {Elahi}}, \bibinfo {author} {\bibfnamefont
  {K.}~\bibnamefont {Artyushkova}}, \bibinfo {author} {\bibfnamefont {T.~E.}\
  \bibnamefont {Beechem}}, \bibinfo {author} {\bibfnamefont {P.~M.}\
  \bibnamefont {Norris}}, \bibinfo {author} {\bibfnamefont {Z.~C.}\
  \bibnamefont {Leseman}}, \bibinfo {author} {\bibfnamefont {J.~T.}\
  \bibnamefont {Gaskins}},\ and\ \bibinfo {author} {\bibfnamefont {P.~E.}\
  \bibnamefont {Hopkins}},\ }\href {https://doi.org/10.1103/PhysRevB.93.140201}
  {\bibfield  {journal} {\bibinfo  {journal} {Physical Review B}\ }\textbf
  {\bibinfo {volume} {93}},\ \bibinfo {pages} {140201(R)} (\bibinfo {year}
  {2016})}\BibitemShut {NoStop}%
\bibitem [{\citenamefont {Pan}\ \emph {et~al.}(2020)\citenamefont {Pan},
  \citenamefont {Zhou},\ and\ \citenamefont {Chen}}]{Pan2020}%
  \BibitemOpen
  \bibfield  {author} {\bibinfo {author} {\bibfnamefont {Y.}~\bibnamefont
  {Pan}}, \bibinfo {author} {\bibfnamefont {J.}~\bibnamefont {Zhou}},\ and\
  \bibinfo {author} {\bibfnamefont {G.}~\bibnamefont {Chen}},\ }\href
  {https://doi.org/10.1103/PhysRevB.101.144203} {\bibfield  {journal} {\bibinfo
   {journal} {Physical Review B}\ }\textbf {\bibinfo {volume} {101}},\ \bibinfo
  {pages} {144203} (\bibinfo {year} {2020})}\BibitemShut {NoStop}%
\end{thebibliography}%

\end{document}